\documentclass{IEEEtran}
\usepackage{amsmath}
\usepackage{amsmath,amsfonts}
\usepackage{amsmath,bm}
\usepackage{algorithmic}
\usepackage{algorithm}
\usepackage{array}
\usepackage[caption=false,font=normalsize,labelfont=sf,textfont=sf]{subfig}
\usepackage{textcomp}
\usepackage{stfloats}
\usepackage{url}
\usepackage{verbatim}
\usepackage{graphicx}
\usepackage{cite}
\usepackage{enumitem}
\usepackage{diagbox}
\usepackage{multirow}
\usepackage{threeparttable}
\usepackage{tabularx}
\usepackage{booktabs}
\usepackage{threeparttable}

\setlist[description]{labelwidth = 3cm, leftmargin = 3.2cm}

\makeatletter
\def\ps@IEEEtitlepagestyle{%
  \def\@oddfoot{\mycopyrightnotice}%
  \def\@evenfoot{}%
}
\def\mycopyrightnotice{%
    {\hfill \footnotesize This work has been submitted to the IEEE for possible publication. Copyright may be transferred without notice, after which this version may no longer be accessible.\hfill}
}
\makeatother

\begin{document}

\title{Real-time computational powered landing guidance using convex optimization and neural networks}

\author{Zhipeng Shen,~
Shiyu Zhou,~
Jianglong Yu
\thanks{This work was supported by the National Natural Science Foundation of China under Grants 62103016, 61922008, the Foundation Strengthening Program Technology Field Fund under Grant 2021-JCJQ-JJ-1237,  China National Postdoctoral Program for Innovative Talents under Grant BX20200034, and the China Postdoctoral Science Foundation under Grant 2020M680297.}
\thanks{Z. Shen is with the Department of Aeronautical and Aviation Engineering, the Hong Kong Polytechnic University, Hong Kong, China (E-mail: zhipeng.shen@connect.polyu.hk). 
}
\thanks{S. Zhou is with the Department of Biomedical Engineering, City University of Hong Kong, Hong Kong, China. 
}
\thanks{J. Yu is with the School of Automation Science and Electrical Engineering, Beihang University, Beijing, China (E-mail: sdjxyjl@buaa.edu.cn).
}}



\maketitle

\begin{abstract}
Computational guidance is an emerging and accelerating trend in aerospace guidance and control. Combining machine learning and convex optimization, this paper presents a real-time computational guidance method for the 6-degrees-of-freedom powered landing guidance problem. The powered landing guidance problem is formulated as an optimal control problem, which is then transformed into a convex optimization problem. Instead of brutally using the neural networks as the controller,  we use neural networks to improve the state-of-the-art sequential convex programming (SCP) algorithm. Based on the deep neural network, an initial trajectory generator is designed to provide a satisfactory initial guess for the SCP algorithm. Benefitting from designing the initial trajectory generator as a sequence model predictor, the proposed data-driven SCP architecture is capable of improving the performance of any state-of-the-art SCP algorithm in various applications, not just powered landing guidance. The simulation results show that the proposed method can precisely guide the vehicle to the landing site. Moreover, through Monte Carlo tests, the proposed method can averagely save 40.8${\bf{\% }}$ of the computation time compared with the SCP method, while ensuring higher terminal states accuracy. The proposed computational guidance scheme is suitable for real-time applications.

\end{abstract}



\section{Introduction}
\IEEEPARstart{I}{n} recent years, computational guidance is an emerging trend in aerospace guidance and control \cite{lu2021,Wang2019,liu2019,kunhippurayil2021,lu2017,liu2017}. Thanks to the improved computing power and advanced algorithms, computational guidance methods can leverage the increased onboard computational capability to complete the intensive online computation \cite{lu2017,lu2021}. However, instead of simply using computing power to numerically solve the guidance problem via brute force, computational guidance achieves online computation through upfront investment in problem formulation, modeling, and analysis \cite{lu2017,liu2019}. Motivated by the computational guidance philosophy \cite{lu2021,Wang2019,lu2017,liu2019,liu2017,kunhippurayil2021}, this paper introduces a data-driven method to further enhance the upfront investment and achieve more efficient online computation.

\begin{figure}[!ht]
\centering
\includegraphics[width=8cm]{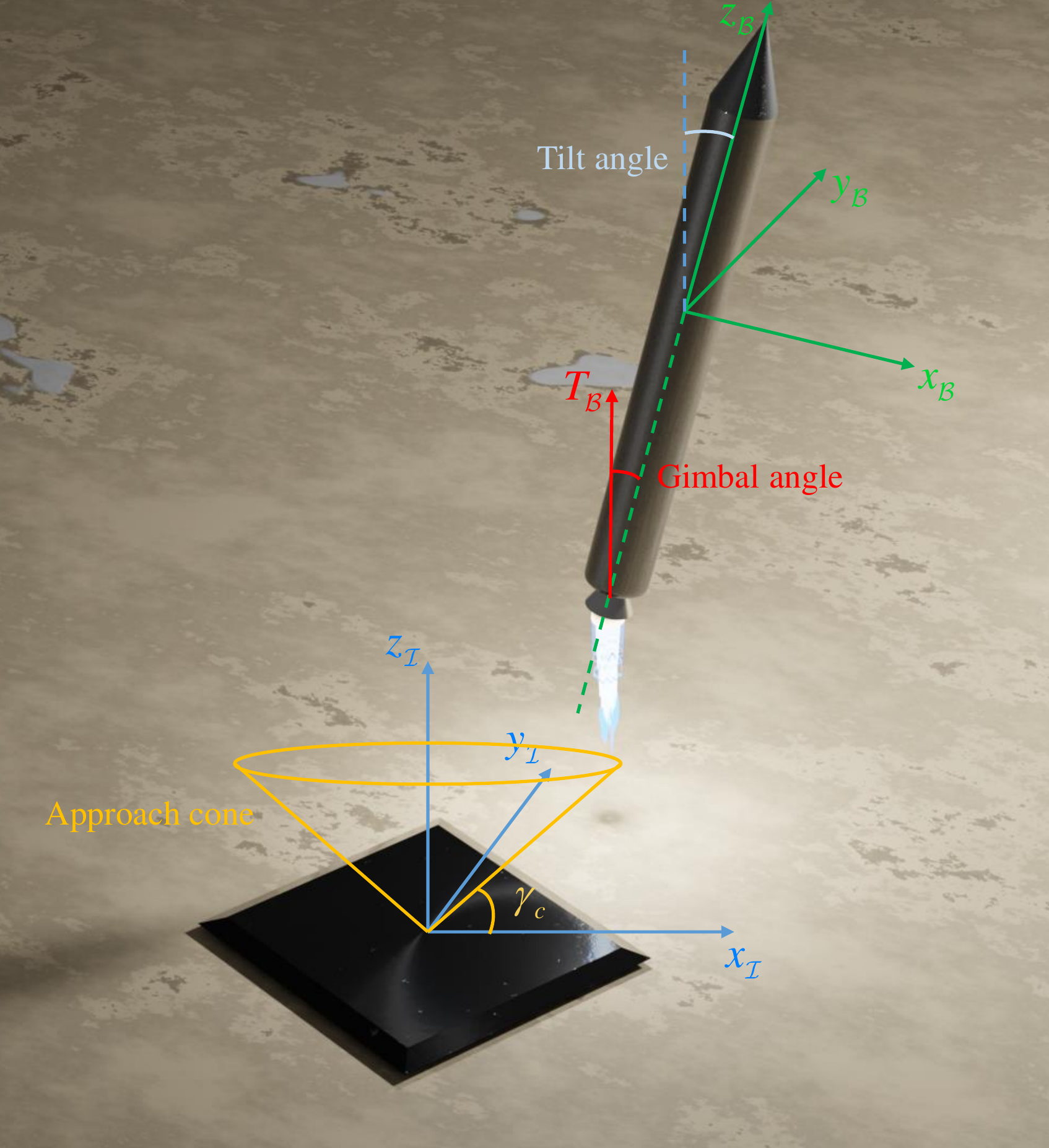}
\caption{The landing scenario.}
\label{fig_scenario}
\end{figure}

As a popular and powerful way, the guidance method based on convex optimization is a significant computational guidance method \cite{kunhippurayil2021,acikmese2011,harris2014,acikmese2013,liu2016a,benedikter2021,liu2016,szmuk2020,reynolds2020,szmuk2018,shen2022,chen2021,malyuta2022convex}. The convex optimization-based guidance method transforms the optimal control problem into a convex optimization problem by convexification and then utilizes the state-of-the-art convex optimization solver to efficiently solve the convex optimization problem. Lossless convexification theory bridges the gap between nonconvex optimal control problems and efficient convex optimization methods \cite{kunhippurayil2021,acikmese2011,harris2014,acikmese2013}. The results of lossless convexification have been extended to optimal control problems with nonconvex control constraints \cite{acikmese2011}, linear state constraints \cite{harris2014}, and annular control constraints \cite{kunhippurayil2021}. For optimal control problems with more general nonconvexities, such as the complex 6-degrees-of-freedom (DoF) dynamics, researchers turn to sequential convex programming (SCP) methods \cite{szmuk2020,reynolds2020,szmuk2018}. Through linearization, SCP solves the original nonconvex optimal control problem by solving a sequence of locally convex approximations. The current researches provide theoretical guarantees for locally optimal solutions obtained by SCP methods with guaranteed convergence properties \cite{mao2016,mao2017,mao2019}. This method has been extended to many applications of aerospace guidance, such as atmospheric reentry \cite{liu2016}, missile guidance \cite{liu2016a}, rocket launch \cite{benedikter2021}, and optimal landing \cite{Wang2019,liu2019,szmuk2018}. Since the linearization based on Taylor expansion can make the SCP method suitable for different nonlinear systems, the SCP method can be applied in many different applications. However, the linearization in the SCP algorithm also makes its performance significantly dependent on the initial reference trajectory. One of the contributions of this paper is to design an initial trajectory generator through the deep neural network (DNN) to improve the quality of the initial reference trajectory and further improve the performance of the SCP method.

The optimal landing problem is one of the optimal control problems that has been widely concerned because it is widely used in Mars exploration missions and reusable rocket missions \cite{Wang2019,liu2019,kunhippurayil2021,szmuk2018}. Since lossless convexification can transform the nonconvex 3-DoF landing problem into a more easily solved and equivalent convex optimization problem, lossless convexification was widely used in optimal landing tasks and has been verified in a series of flight experiments \cite{kunhippurayil2021,acikmese2013,scharf2014,scharf2017,dueri2017}. More recently, researchers begin to consider more general optimal landing problems with 6-DoF dynamics \cite{szmuk2020,reynolds2020,szmuk2018,chai2020}. In \cite{szmuk2020}, an SCP algorithm was presented for a generalized 6-DoF free-final-time powered descent guidance problem while considering the state-triggered constraints. An SCP algorithm was also introduced for the 6-DoF powered descent guidance problem with dual quaternion-based dynamics in \cite{reynolds2020}. To achieve more efficient computational guidance methods, this paper presents a data-driven SCP approach by combining SCP with DNN. The proposed approach can significantly reduce the number of iterations required for the convergence of SCP methods and further improve the computational performance.

Recently, researchers are also interested in new techniques for onboard algorithms leveraging advancements in machine learning \cite{shi2021,cheng2020,cheng2021,schiassi2022,you2021,li2022a,chai2020a,chai2020,chai2022design,ma2022local,yu2018practical,dong2022neuroadaptive}. Many researchers have combined traditional methods with neural networks to improve computational performance \cite{shi2021,cheng2020,cheng2021,schiassi2022,you2021,li2022a,chai2020a,chai2020,chai2022design}. Shi et al. in \cite{shi2021} proposed a method based on deep learning to realize real-time online trajectory planning of hypersonic vehicles. Chen et al. used DNN to generate the optimal asteroid landing trajectory \cite{cheng2020}. Cheng et al. used DNN to improve the predictor-corrector entry guidance law of lifting hypersonic vehicles \cite{cheng2021}. The optimal planetary orbit transfer was studied by combining physics-informed neural networks and Pontryagin's minimum principle in \cite{schiassi2022}. You et al. in \cite{you2021} studied the fuel-optimal powered decent guidance by incorporating Pontryagin's minimum principle with neural network. In \cite{li2022a}, the free final-time fuel-optimal powered landing guidance problem was studied by combining the lossless convex optimization and the DNN, where the DNN was used to predict the final time. Chai et al. developed a real-time optimal command generator for reentry problems by using the DNN in \cite{chai2020a} and further proposed a DNN-based method to achieve the integration of trajectory planning and attitude control for 6-DOF reentry flight \cite{chai2020}. Since neural networks are usually considered black boxes, it is not a guaranteed method to directly use them as control command generators. Instead of using neural networks to output control commands, this paper combines DNN with SCP methods. In this paper, a DNN-based trajectory generator is proposed to generate initial guesses required by the SCP method. This method not only improves the performance of the SCP method through DNN but also provides the same theoretical guarantee as the SCP method. Similarly, a warm-start method was proposed in \cite{banerjee2020b}, which used neural networks to predict the initial trajectory for the SCP method. However, the neural network in \cite{banerjee2020b} was used to predict a trajectory parameterized as a polynomial for each state and control. In this case, it is difficult to determine the order of each polynomial in different applications. In \cite{kim2022guided}, a neural network policy was updated through the trajectories obtained from a one-step SCP iteration and a feedback control law. The control law in \cite{kim2022guided} needs to be carefully designed according to the specific application. Inspired by the sequence model in natural language processing \cite{zhang2021dive,goodfellow2016deep}, we design the initial trajectory generator as a sequence model predictor. Therefore, the initial trajectory generator does not need to be specially designed for the problem. Since the trajectories of different problems can be considered as the same sequence model with different dimensions on each discrete point, this makes the proposed architecture applicable to SCP algorithms in various applications.

The primary contributions of this paper are highlighted as follows.
\begin{itemize}
  \item [1)] 
  A real-time computational guidance approach is proposed for the powered landing guidance. Compared with the state-of-the-art SCP algorithm \cite{szmuk2020,reynolds2020,szmuk2018}, the proposed algorithm saves 40.8$\%$ time and has better accuracy. The proposed method can meet real-time requirements better.
      
  \item [2)]
  Instead of brutally using the neural networks to output optimal solutions \cite{chai2020a,chai2020,chai2022design}, a data-driven SCP algorithm is presented by combining SCP and DNN, such that the proposed method can improve the performance while providing the same theoretical guarantee as the SCP algorithm.
  
  \item [3)]
  The initial trajectory generator is designed as a sequence model predictor, which makes the proposed computational guidance architecture a more general approach that can be applied to improve the computational performance of any SCP algorithm in various applications. Meanwhile, the related works \cite{banerjee2020b,kim2022guided} requires special design according to the specific applications.
\end{itemize}

The remainder of this paper is organized as follows. Sec. \ref{Landing problem} shows the formulation of the 6-DoF landing problem. Sec. \ref{Convex programming} introduces the SCP algorithm. Sec. \ref{Initial Trajectory Generator} gives the details of the proposed initial trajectory generator, along with the proposed SCP algorithm. Sec. \ref{Results} presents the test results of the proposed guidance approach. Sec. \ref{Conclusion} concludes the whole paper.

\section{Problem Statement}\label{Landing problem}
In this section, the free-final-time 6DoF powered landing guidance problem is formulated as a nonconvex optimal control problem, while considering the aerodynamic effects and the multiple constraints. The landing scenario is shown in Fig. \ref{fig_scenario}.

\subsection{Notation}
We use $ \cdot $ to denote the vector dot product, $ \times $ to represent the vector cross product, and $\left\|  \cdot  \right\|$ to denote the Euclidean norm. We denote time as $t \in \mathbb{R}$. The initial time ${t_0}$ is defined as the time at which the guidance problem begins, and the terminal time ${t_f}$ is defined as the time at which the vehicle reaches the terminal conditions. Subscripts $\mathcal{I}$ and $\mathcal{B}$ represent parameters expressed in the inertial frame $\mathcal{F_I}$ and body-fixed frame $\mathcal{F_B}$, respectively. The frame $\mathcal{F_I}$ is an east-north-up coordinate frame, and its origin is fixed at the landing site. The origin of the frame $\mathcal{F_B}$ coincides with the vehicle’s center of mass, the $z$ axis of $\mathcal{F_B}$ points along the vertical axis of the vehicle, the $x$ axis of $\mathcal{F_B}$ points out the side of the vehicle, and the $y$ axis of $\mathcal{F_B}$ completes the right-handed system.
Some notations associated with the guidance problem are defined below.
\subsubsection*{\bf Notations}

\begin{description}{}{}
\item[${g_0}$] Standard gravitational acceleration
\item[$\rho $] Ambient atmospheric density
\item[$m(t) \in {\mathbb{R}_{ +  + }}$] Vehicle mass
\item[${\bm{r}_\mathcal{I}}(t) \in {\mathbb{R}^3}$] Inertial position of the vehicle
\item[${\bm{v}_\mathcal{I}}(t) \in {\mathbb{R}^3}$] Inertial velocity of the vehicle
\item[${\bm{g}_\mathcal{I}} \in {\mathbb{R}^3}$] Gravitational acceleration
\item[${\bm{T}_\mathcal{I}}(t) \in {\mathbb{R}^3}$] Thrust vector
\item[${\bm{A}_\mathcal{I}}(t) \in {\mathbb{R}^3}$] Aerodynamic force
\item[${S_A}\in\mathbb{R}_{ +  + }$] Reference area of the vehicle
\item[${C_A} \in \mathbb{S}_{ +  + }^3$] Aerodynamic coefficient matrix
\item[$ \otimes $] Quaternion multiplication
\item[${\bm{q}_{i}}$] Identity quaternion
\item[${\bm{q}_{{\mathcal{I}}  {\mathcal{B}}}}(t) \in {{\mathcal{S}}^3} \subset {\mathbb{R}^4}$] Unit quaternion that parameterizes the transformation from $\mathcal{F_I}$ to $\mathcal{F_B}$
\item[${C_{{{\cal I}{\cal B}}}}(t) \in {\rm{SO}}(3)$]Direction cosine matrix corresponding to ${q_{{\mathcal{I}}  {\mathcal{B}}}}(t)$, ${C_{{{\cal I}{\cal B}}}}(t) \buildrel \Delta \over = {C_{{{\cal I}{\cal B}}}}\left( {{{\bm{q}}_{{{\cal I}{\cal B}}}}(t)} \right)$
\item[${\bm{\omega} _{{\cal B}}}(t) \in {\mathbb{R}^3}$] Angular velocity of the vehicle in $\mathcal{F_B}$
\item[${J_{{\cal B}}} \in \mathbb{S}_{ +  + }^3$]Moment of inertia of the vehicle
\item[$\bm{d}_{T, \cal{B}}$] Position of the engine gimbal pivot point in $\mathcal{F_B}$
\item[$\bm{d}_{A, \cal{B}}$] Position of the center of pressure in $\mathcal{F_B}$

\item[$\bm{M}_{\cal{B}}(t)\in \mathbb{R}^3$] Torque acting on the vehicle, $\bm{M}_{\cal{B}}(t) \buildrel \Delta \over = \bm{d}_{T, \cal{B}} \times \bm{T}_{\cal{B}}(t)+\bm{d}_{A, \cal{B}} \times \bm{A}_{\cal{B}}(t) $
\item[${I_{{\rm{sp}}}}$] Vacuum specific impulse of the engine
\item[${P_{{\rm{atm}}}}$] Ambient atmospheric pressure
\item[${S_{{\rm{ne}}}}$] Nozzle exit area of the engine
\end{description}
\subsection{Dynamics}
Since most powered landing maneuvers are far below the orbital velocities, and the initial position of the powered landing is only a few kilometers away from the landing site. The effects of planetary rotation are neglected, and we assume the gravitational field is uniform. Moreover, higher-order phenomena such as fuel slosh and elastic structural modes are not considered. As a rigid body, the vehicle has a constant center of mass and moment of inertia. Additionally, the density and pressure of the ambient atmosphere are assumed to be constant, and the effects of winds are not taken into account. We also assume that the center of mass is fixed to $\mathcal{F_B}$. In this paper, the attitude dynamics are established based on quaternion, and the scalar-first quaternion convention is used.

The mass-depletion dynamics are given by
\begin{equation}
    \label{eq1}
    \dot m(t) =  - \alpha {\left\| {{{\bm{T}}_\mathcal{B}}(t)} \right\|} - \beta 
\end{equation}
where ${\alpha} \buildrel \Delta \over = 1/({I_{{\rm{sp}}}}{g_0})$ and ${\beta}\buildrel \Delta \over= {\alpha}{P_{{\rm{atm}}}}{S_{{\rm{ne}}}}$.

The translational dynamics are given as follows
\begin{equation}
    \label{eq2}
    {\dot{\bm{r}}_{\cal I}}(t) = {{\bm{v}}_{\cal I}}(t)
\end{equation}
\begin{equation}
    \label{eq3}
    {{{\bm{\dot v}}}_{\cal I}}(t) = \frac{1}{{m(t)}}({C_{{\cal B}{\cal I}}}(t){{\bm{T}}_{\cal B}}(t) + {{\bm{A}}_{\cal I}}(t)) + {{\bm{g}}_{\cal I}}
\end{equation}
where ${C_{{\cal B}{\cal I}}}(t) \buildrel \Delta \over = C_{{\cal I}{\cal B}}^T(t) = {C_{{\cal I}{\cal B}}}\left( {q_{{\cal I}{\cal B}}^ * (t)} \right)$, and ${q_{{\cal I}{\cal B}}^ * (t)}$ is the conjugate of ${q_{{\cal I}{\cal B}}}(t)$. The aerodynamic force ${\bm{A}_\mathcal{I}}(t)$ is modeled as follows
\begin{equation}
    \label{eq4}
    {\bm{A}_{\cal I}}(t) =  - \frac{1}{2}\rho {\left\| {{\bm{v}_{\cal I}}(t)} \right\|}{S_A}{C_A}{\bm{v}_{\cal I}}(t)
\end{equation}
where ${C_A}$ is a diagonal matrix for most vehicles that are approximately axisymmetric. The attitude dynamics are given by
\begin{equation}
    \label{eq5}
    {{{\bm{\dot q}}}_{{\cal I}{\cal B}}}(t) = \frac{1}{2}\bm{\Omega} \left( {{\bm{\omega} _{\cal B}}(t)} \right){{\bm{q}}_{{\cal I} {\cal B}}}(t)
\end{equation}
\begin{equation}
    \label{eq6}
    {J_{\cal B}}{{\dot {\bm{\omega}} }_{\cal B}}(t) = {{\bm{M}}_{\cal B}}(t) - {{\bm{\omega}} _{\cal B}}(t) \times {J_{\cal B}}{{\bm{\omega}} _{\cal B}}(t)
\end{equation}
where $\bm{\Omega} \left(  \cdot  \right)$ is a skew-symmetric matrix defined for the quaternion kinematics (\ref{eq5}).

\subsection{State Constraints}
Since the fuel on the vehicle is limited, the mass of the vehicle should be greater than a minimum mass ${m_{{\rm{min}}}} \in {\mathbb{R}_{ +  + }}$. The mass is constrained by enforcing
\begin{equation}
    \label{eq7}
    {m_{{\rm{min}}}} \le m(t)
\end{equation}

Next, the trajectory of the vehicle is constrained in an approach cone to ensure that the vehicle has sufficient elevation on the planetary surface for landing. The approach cone constraint can be expressed as
\begin{equation}
    \label{eq8}
    {\left\| {{H_c}{{\bm{r}}_{{\cal I}}}(t)} \right\|} \le \cot {\gamma _c}{\bm{e}} \cdot {{\bm{r}}_{{\cal I}}}(t)
\end{equation}
\begin{equation}
    \label{eq9}
    {\bm{e}} \buildrel \Delta \over = {\left[ {\begin{array}{*{20}{c}}
0&0&1
\end{array}} \right]^T}, {H_c} \buildrel \Delta \over = \left[ {\begin{array}{*{20}{c}}
1&0&0\\
0&1&0
\end{array}} \right]
\end{equation}
The approach cone is centered at the landing point, and ${\gamma _c} \in \left[ {{{0^\circ }},{{90}^\circ }} \right)$ is the angle between the cone and the horizontal.

The vehicle's tile angle, which is the angle between the ${z}$ axes of ${\cal{F}}_{\cal{B}}$ and ${\cal{F}}_{\cal{I}}$, should also be constrained by enforcing
\begin{equation}
    \label{eq10}
    2{\left\| {{H_t}{{\bm{q}}_{{{\cal I}{\cal B}}}}(t)} \right\|^2} \le 1 - \cos {\theta _{\max }}
\end{equation}
\begin{equation}
    \label{eq11}
    {H_t} \buildrel \Delta \over = \left[ {\begin{array}{*{20}{c}}
0&1&0&0\\
0&0&1&0
\end{array}} \right]
\end{equation}
where ${\theta _{\max }} \in \left( {{0^\circ },{{90}^\circ }} \right]$ is the maximum tilt angle.

The allowable angular velocity is constrained by enforcing
\begin{equation}
    \label{eq12}
    {\left\| {{{\bm{\omega }}_{{\cal B}}}(t)} \right\|_\infty} \le {\omega _{\max }}
\end{equation}
where ${\omega _{\max }} \in {\mathbb{R}_{ +  + }}$ is a maximum angular velocity, and ${\left\|  \cdot  \right\|_\infty }$ denotes the $\infty$-norm.

\subsection{Control Constraints}
We assume that a single gimbaled engine is equipped on the vehicle. The engine can rotate symmetrically, but is constrained to a maximum gimbal angle ${\vartheta _{\max }} \in ({0^ \circ },{90^ \circ })$. Hence, the first control constraint is given by
\begin{equation}
    \label{eq13}
    {\left\|  {H_c}{{{\bm{T}}_{{\cal B}}}(t)} \right\|} \le \tan {\vartheta _{\max }}{\bm{e}} \cdot {{\bm{T}}_{{\cal B}}}(t)
\end{equation}

The thrust magnitude of the engine can vary between a fixed minimum value and a fixed maximum value. The second control constraint is given as
\begin{equation}
    \label{eq14}
    T_{\min } \leq\left\|\boldsymbol{T}_{\mathcal{B}}(t)\right\| \leq T_{\max }
\end{equation}
where $\left[ {{T_{\min }},{T_{\max }}} \right] \subset {\mathbb{R}_{ +  + }}$ is the permitted thrust interval. We assume that the rocket engine starts at the initial time $t_0$, and only the minimum thrust magnitude with a zero-degree gimbal angle can be used when the engine starts. Then, an equality constraint can be given as
\begin{equation}
    \label{T0}
    {{\bm{T}}_{\cal{B}}}\left( {{t_0}} \right) ={{\bm{T}}_{{\cal{B}}\rm{0}}}= {\left[ {\begin{array}{*{20}{c}}
0&0&{{T_{\min }}}
\end{array}} \right]^T}
\end{equation}

\subsection{Nonconvex Optimal Control Problem} \label{2E}
We conclude this section by completing the statement of the nonconvex optimal control problem. The objective function and boundary conditions of the optimal control problem can be designed according to the scenario and mission requirements. In this work, we mainly focus on the minimum-fuel problem, which is equivalent to maximizing the terminal mass. The initial conditions can be given as
\begin{equation}
    \label{eq15}
    m\left( {{t_0}} \right) = {m_0},{{\bm{r}}_{{\cal I}}}\left( {{t_0}} \right) = {{\bm{r}}_{{{\cal I}}{\rm{0}}}}, {{\bm{v}}_{{\cal I}}}\left( {{t_0}} \right) = {{\bm{v}}_{{{\cal I}}{\rm{0}}}}
\end{equation}
\begin{equation}
    \label{eq16}
    {{\bm{q}}_{{{\cal I}{\cal B}}}}\left( {{t_0}} \right) = {{\bm{q}}_0},{{\bm{\omega }}_{{\cal B}}}\left( {{t_0}} \right) = {{\bm{\omega }}_{{{\cal B}}{\rm{0}}}}
\end{equation}
where $m_0$, ${{\bm{r}}_{{{\cal I}}{\rm{0}}}}$, ${{\bm{v}}_{{{\cal I}}{\rm{0}}}}$, ${{\bm{q}}_0}$, ${{\bm{\omega }}_{{{\cal B}}{\rm{0}}}}$ are the prescribed mass, position, velocity, quaternion, and angular velocity at the initial time, respectively. At the terminal time, the goal is to land the vehicle at the landing site steadily and safely. The terminal conditions are given by
\begin{equation}
    \label{eq17}
    {{\bm{r}}_{{\cal I}}}\left( {{t_f}} \right) = \bm{0}, {{\bm{v}}_{{\cal I}}}\left( {{t_f}} \right) = \bm{0},{{\bm{q}}_{{{\cal I}{\cal B}}}}\left( {{t_f}} \right) = {{\bm{q}}_i},{{\bm{\omega }}_{{\cal B}}}\left( {{t_f}} \right) = \bm{0}
\end{equation}
The nonconvex optimal control problem is summarized as Problem 1.

\begin{table*}[!ht]
\centering 

\begin{tabular}{p{3cm}<{\raggedright}cccc} \toprule[1pt] 
\multicolumn{5}{p{15cm}<{\raggedright}}{\emph{Problem 1}: Find the control commands profile ${\bm{T}}_{\cal{B}}\left(t\right), \forall t \in\left[t_0, t_f\right]$ to solve the nonconvex optimal control problem.}  \\ \hline
\emph{Cost function} &\multicolumn{2}{p{6cm}<{\raggedright}}{$\mathop {\min }\limits_{{t_f},{{\bm{T}}_{{\cal B}}}(t)} \quad  - m\left( {{t_f}} \right)$} & & Convex   \\ \hline

\multicolumn{1}{l}{\multirow{3}{*}{\emph{Boundary conditions}}}&\multicolumn{2}{p{7cm}<{\raggedright}}{}&&\\
&	\multicolumn{2}{p{7cm}<{\raggedright}}{$m\left( {{t_0}} \right) = {m_0},{{\bm{r}}_{{\cal I}}}\left( {{t_0}} \right) = {{\bm{r}}_{{{\cal I}}{\rm{0}}}}, {{\bm{v}}_{{\cal I}}}\left( {{t_0}} \right) = {{\bm{v}}_{{{\cal I}}{\rm{0}}}}$ ${{\bm{q}}_{{{\cal I}{\cal B}}}}\left( {{t_0}} \right) = {{\bm{q}}_0},{{\bm{\omega }}_{{\cal B}}}\left( {{t_0}} \right) = {{\bm{\omega }}_{{{\cal B}}{\rm{0}}}},{{\bm{T}}_{\cal{B}}}\left( {{t_0}} \right) ={{\bm{T}}_{{\cal{B}}\rm{0}}}$} &	See Eqs. (\ref{T0}-\ref{eq17}) & Convex \\ 
& \multicolumn{2}{p{7cm}<{\raggedright}}{${{\bm{r}}_{{\cal I}}}\left( {{t_f}} \right) = \bm{0}, {{\bm{v}}_{{\cal I}}}\left( {{t_f}} \right) = \bm{0},{{\bm{q}}_{{{\cal I}{\cal B}}}}\left( {{t_f}} \right) = {{\bm{q}}_i},{{\bm{\omega }}_{{\cal B}}}\left( {{t_f}} \right) = \bm{0}$} & & \\ \hline

\multicolumn{1}{l}{\multirow{3}{*}{\emph{Dynamics}}}&\multicolumn{2}{p{7cm}<{\raggedright}}{}&&\\
&	\multicolumn{2}{p{7cm}<{\raggedright}}{$\dot m(t) =  - \alpha {\left\| {{{\bm{T}}_\mathcal{B}}(t)} \right\|} - \beta$} &	See Eq. (\ref{eq1}) & Nonconvex \\ 
& \multicolumn{2}{p{7cm}<{\raggedright}}{${\dot{\bm{r}}_{\cal I}}(t) = {{\bm{v}}_{\cal I}}(t)$} &	See Eq. (\ref{eq2}) & Nonconvex \\ 
&\multicolumn{2}{p{7cm}<{\raggedright}}{${{{\bm{\dot v}}}_{\cal I}}(t) = \frac{1}{{m(t)}}({C_{{\cal B}{\cal I}}}(t){{\bm{T}}_{\cal B}}(t) + {{\bm{A}}_{\cal I}}(t)) + {{\bm{g}}_{\cal I}}$} &	See Eq. (\ref{eq3}) &Nonconvex \\ 
&\multicolumn{2}{p{7cm}<{\raggedright}}{${{{\bm{\dot q}}}_{{\cal I}{\cal B}}}(t) = \frac{1}{2}\bm{\Omega} \left( {{\bm{\omega} _{\cal B}}(t)} \right){{\bm{q}}_{{\cal I} {\cal B}}}(t)$} &	See Eq. (\ref{eq5}) & Nonconvex \\ 
&\multicolumn{2}{p{7cm}<{\raggedright}}{${J_{\cal B}}{{\dot {\bm{\omega}} }_{\cal B}}(t) = {{\bm{M}}_{\cal B}}(t) - {{\bm{\omega}} _{\cal B}}(t) \times {J_{\cal B}}{{\bm{\omega}} _{\cal B}}(t)$} &	See Eq. (\ref{eq6}) & Nonconvex \\ \hline

\multicolumn{1}{l}{\multirow{3}{*}{\emph{State constraints}}}&\multicolumn{2}{p{7cm}<{\raggedright}}{}&&\\
&	\multicolumn{2}{p{7cm}<{\raggedright}}{${m_{{\rm{min}}}} \le m(t)$} &	See Inequality (\ref{eq7}) & Convex \\   
& \multicolumn{2}{p{7cm}<{\raggedright}}{${\left\| {{H_c}{{\bm{r}}_{{\cal I}}}(t)} \right\|} \le \cot {\gamma _c}{\bm{e}} \cdot {{\bm{r}}_{{\cal I}}}(t)$} &	See Inequality (\ref{eq8}) & Convex \\ 
&\multicolumn{2}{p{7cm}<{\raggedright}}{$2{\left\| {{H_t}{{\bm{q}}_{{{\cal I}{\cal B}}}}(t)} \right\|^2} \le 1 - \cos {\theta _{\max }}$} &	See Inequality (\ref{eq10}) & Convex\\ 
&\multicolumn{2}{p{7cm}<{\raggedright}}{${\left\| {{{\bm{\omega }}_{{\cal B}}}(t)} \right\|}_\infty \le {\omega _{\max }}$} &	See Inequality (\ref{eq12}) & Convex \\ \hline

\multicolumn{1}{l}{\multirow{3}{*}{\emph{Control constraints}}}&\multicolumn{2}{p{7cm}<{\raggedright}}{}&&\\
&	\multicolumn{2}{p{7cm}<{\raggedright}}{${\left\|  {H_c}{{{\bm{T}}_{{\cal B}}}(t)} \right\|} \le \tan {\vartheta _{\max }}{\bm{e}} \cdot {{\bm{T}}_{{\cal B}}}(t)$} &	See Inequality (\ref{eq13}) & Convex \\   
&\multicolumn{2}{p{7cm}<{\raggedright}}{$T_{\min } \leq\left\|\boldsymbol{T}_{\mathcal{B}}(t)\right\| \leq T_{\max }$} &	See Inequality (\ref{eq14}) & Nonconvex \\ \bottomrule[1pt] 

\end{tabular}
\end{table*}

\section{Sequential Convex Programming} \label{Convex programming}
This section introduces a SCP algorithm to solve Problem 1. In Sec. \ref{Convex subproblem}, Problem 1 is converted into a discrete-time convex optimization subproblem. In Sec. \ref{SCP algorithm}, the SCP algorithm is presented, which iteratively solves a sequence of subproblems to get a converged solution.

\subsection{Convex Formulation} \label{Convex subproblem}
\emph{1. \quad Normalization} 

The dynamics equation in Problem 1 is a nonconvex factor. Thus, it should be converted into a convex formulation. The continuous-time dynamics can be represented as
\begin{equation}
    \label{dynamics in f}
    {\bm{\dot x}}(t) = f({\bm{x}}(t),{\bm{u}}(t)),\quad \forall t \in \left[ {{t_{\rm{0}}},{t_f}} \right]
\end{equation}
\begin{subequations}
\begin{gather}
{\bm{x}}(t)  \buildrel \Delta \over  = {\left[ {\begin{array}{*{20}{l}}
{m(t)}&{{\bm{r}}_{{\cal I}}^T(t)}&{{\bm{v}}_{{\cal I}}^T(t)}&{{\bm{q}}_{{{\cal I}{\cal B}}}^T(t)}&{{\bm{\omega }}_{{\cal B}}^T(t)}
\end{array}} \right]^T}\\
{\bm{u}}(t)  \buildrel \Delta \over  = {{\bm{T}}_{{\cal B}}}(t)
\end{gather}
\end{subequations}
where ${\bm{x}}(t)  \in {\mathbb{R}^{{n_x}}}$ and ${\bm{u}}(t) \in {\mathbb{R}^{{n_u}}}$ denote the state and control vectors, respectively, and $f:{\mathbb{R}^{{n_x}}} \times {\mathbb{R}^{{n_u}}} \to {\mathbb{R}^{{n_x}}}$ denotes the continuous-time nonconvex dynamics. To begin, the scaled time $\tau \in \left[ 0,1 \right]$ is defined to equivalently convert Problem 1 into a fixed-final-time problem. By applying the chain rule, the dynamics can be rewritten as
\begin{equation}
    \label{time scale 1}
    {\bm{x}^{\prime}}(\tau )\buildrel \Delta \over = \frac{d}{{d\tau }}{\bm{x}}(\tau ) = \frac{{dt}}{{d\tau }}f({\bm{x}}(\tau ),{\bm{u}}(\tau ))
\end{equation}
We assume that ${t_{\rm{0}}} = 0$. Thus, one has $t = {t_f}\tau $. Further, the Eq. (\ref{time scale 1}) can be written as
\begin{equation}
    \label{time scale 2}
    \boldsymbol{x}^{\prime}(\tau)=t_f  f(\boldsymbol{x}(\tau), \boldsymbol{u}(\tau))\buildrel \Delta \over =F(\boldsymbol{x}(\tau), \boldsymbol{u}(\tau), t_f)
\end{equation}

\emph{2. \quad Linearization} 

To formulate a convex problem, the equality constraint functions must be affine \cite{boyd2004convex}. Hence, Eq. (\ref{time scale 2}) is linearized as follows
\begin{equation}
    \label{Linearization}
    {\bm{x}^{\prime}}(\tau ) \approx A(\tau ){\bm{x}}(\tau ) + B(\tau ){\bm{u}}(\tau ) + {\bm{s}}(\tau )t_f + {\bm{c}}(\tau )
\end{equation}
\begin{equation}
    \label{Partial differential}
    A(\tau ) \buildrel \Delta \over = {\left. {\frac{{\partial F}}{{\partial {\bm{x}}}}} \right|_{\tilde z(\tau )}},B(\tau ) \buildrel \Delta \over = {\left. {\frac{{\partial F}}{{\partial {\bm{u}}}}} \right|_{\tilde z(\tau )}},{\bm{s}}(\tau ) \buildrel \Delta \over = {\left. {\frac{{\partial F}}{{\partial t_f}}} \right|_{\tilde z(\tau )}}
\end{equation}
\begin{equation}
    \label{Linear c}
    {\bm{c}}(\tau ) \buildrel \Delta \over =  - A(\tau )\tilde {\bm{x}} (\tau ) - B(\tau )\tilde {\bm{u}} (\tau )
\end{equation}
where ${\tilde{\bm z}}(\tau )\buildrel \Delta \over = {\left[ {\begin{array}{*{20}{c}}
{\tilde{t}_f}&{{{{\tilde{\bm x}}}^T}(\tau )}&{{{{\tilde{\bm u}}}^T}(\tau )}
\end{array}} \right]^T}$ is the reference trajectory.

The other nonconvex factor in Problem 1 is the thrust magnitude lower bound constraint, which can be given as
\begin{equation}
    \label{Lower bound}
    {h_u}({\bm{u}}(\tau )) \buildrel \Delta \over = {T_{\min }} - \left\| {{\bm{u}}(\tau )} \right\| \le 0
\end{equation}
Further, Eq. (\ref{Lower bound}) can be approximated using a first-order Taylor series:
\begin{equation}
\begin{aligned}
    \label{Linear lower bound}
    {h_u}({\bm{u}}(\tau )) & \approx {h_u}(\tilde{\bm{ u}}(\tau )) + {\left. {\frac{{d{h_u}({\bm{u}}(\tau ))}}{{d{\bm{u}}(\tau )}}} \right|_{\tilde{\bm{ u}}(\tau )}}({\bm{u}}(\tau ) - \tilde{\bm{ u}}(\tau ))\\
    & = {T_{\min }} - \frac{{{{\tilde{\bm{ u}}}^T}(\tau )}}{{\left\| {\tilde{\bm{ u}}(\tau )} \right\|}}{\bm{u}}(\tau ) = {T_{\min }} - {H_{\tilde u}}(\tau){\bm{u}}(\tau )
\end{aligned}
\end{equation}

\emph{3. \quad Discretization} 

The optimal control problem is a continuous-time problem, which leads to infinite dimensional space. By introducing $N$ evenly spaced temporal nodes, the problem is discretized into $N-1$ subintervals. The control command is approximated by the affine interpolation given as
\begin{equation}
    \label{Interpolation}
    {\bm{u}}(\tau ) = {{\hat \eta }_k}(\tau ){{\bm{u}}_k} + {\eta _k}(\tau ){{\bm{u}}_{k + 1}},\quad \forall \tau  \in \left[ {{\tau _k},{\tau _{k + 1}}} \right]
\end{equation}
\begin{equation}
    \label{eta}
    {{\hat \eta }_k}(\tau ) = \frac{{{\tau _{k + 1}} - \tau }}{{{\tau _{k + 1}} - {\tau _k}}},\quad {\eta _k}(\tau ) = \frac{{\tau  - {\tau _k}}}{{{\tau _{k + 1}} - {\tau _k}}}
\end{equation}
where $k \in \{ 1,2, \ldots ,N-1\} $, ${\tau _k} = (k - 1)/(N - 1)$, and ${{\bm{u}}_k} \buildrel \Delta \over = {\bm{u}}({\tau _k})$.
Then, the dynamics Eq. (\ref{Linearization}) can be shown as follows for each time interval.
\begin{equation}
\label{eq28}
\begin{aligned}
    {\bm{x}^{\prime}}(\tau ) = & A(\tau ){\bm{x}}(\tau ) + {{\hat \eta }_k}(\tau )B(\tau ){{\bm{u}}_k} + {\eta _k}(\tau )B(\tau ){{\bm{u}}_{k + 1}}\\
    & + {\bm{s}}(\tau )t_f + {\bm{c}}(\tau ), \quad \forall \tau  \in \left[ {{\tau _k},{\tau _{k + 1}}} \right]
\end{aligned}
\end{equation}

The state transition matrix $\Phi \left( {\tau ,{\tau _k}} \right),\forall \tau  \in \left[ {{\tau _k},{\tau _{k + 1}}} \right]$ associated with Eq. (\ref{eq28}) with zero input is given by
\begin{equation}
    \label{Phi}
    \Phi \left( {\tau ,{\tau _k}} \right) = I + \int_{{\tau _k}}^\tau  A (\zeta )\Phi \left( {\zeta ,{\tau _k}} \right){\rm{d}}\zeta 
\end{equation}
where $I$ is an identity matrix with appropriate dimension. The discrete-time state vector can be denoted as ${{\bm{x}}_k} \buildrel \Delta \over = {\bm{x}}({\tau _k})$. By applying the inverse and transitive properties of the state transition matrix, the discrete-time linearized dynamics can be presented as
\begin{equation}
    \label{New dynamics}
    {{\bm{x}}_{k + 1}} = {A_k}{{\bm{x}}_k} + {{\hat B}_k}{{\bm{u}}_k} + {B_k}{{\bm{u}}_{k + 1}} + {\bm{s}_k}{t_f} + {\bm{c}_k}
\end{equation}
where
\begin{subequations}
\label{Propagation}
\begin{align}
&A_k \buildrel \Delta \over =\Phi\left(\tau_{k+1}, \tau_k\right) \label{31a}\\
&{{\hat B}_k}\buildrel \Delta \over=A_k \int_{\tau_k}^{\tau_{k+1}} \Phi^{-1}\left(\tau, \tau_i\right) \hat{\eta}(\tau) B(\tau) \mathrm{d} \tau \\
&{B_k}\buildrel \Delta \over=A_k \int_{\tau_k}^{\tau_{k+1}} \Phi^{-1}\left(\tau, \tau_k\right) {\eta}(\tau) B(\tau) \mathrm{d} \tau \\
&{\bm{s}_k}\buildrel \Delta \over=A_k \int_{\tau_k}^{\tau_{k+1}} \Phi^{-1}\left(\tau, \tau_k\right) \bm{s}(\tau) \mathrm{d} \tau \\
&{\bm{c}_k}\buildrel \Delta \over=A_k \int_{\tau_k}^{\tau_{k+1}} \Phi^{-1}\left(\tau, \tau_i\right) \bm{c}(\tau) \mathrm{d} \tau \label{31e}
\end{align}
\end{subequations}

The propagation method described in Eqs. (\ref{Phi}-\ref{31e}) is analogous to a multiple shooting method, which can improve the convergence performance of the algorithm \cite{szmuk2020,reynolds2020}. Note that Eq. (\ref{Phi}) and Eqs. (\ref{31a}-\ref{31e}) are computed only based on ${\tilde{\bm z}}(\tau )$, the propagation method can be computed simultaneously in implementation, which is computationally efficient.

\emph{4. \quad Trust Region and Virtual Control}

The SCP algorithm is an iterative algorithm based on linearization. Thus, The trust-region constraint is introduced to ensure that the optimized trajectory lies in a region where the linearization is valid. The trust-region constraint is expressed in a quadratic form as
\begin{equation}
    \label{Trust-region constraint}
    {\left\| {{{\bm{x}}_k} - {{\tilde{\bm{ x}}}_k}} \right\|^2} + {\left\| {{{\bm{u}}_k} - {{\tilde{\bm{ u}}}_k}} \right\|^2} \le {\sigma _k}
\end{equation}
where $k \in \{ 1,2, \ldots ,N\} $, and $\sigma _k$ is a trust-region radius. To allow the optimization process to select the trust region, the cost function is augmented with the trust-region radii ${\bm{\sigma }} \in {\mathbb{R}^N_+}$ as
\begin{equation}
    \label{Jtr}
    {J_{{\rm{tr}}}} = {w}_{{\rm{tr}}}{\left\| {\bm{\sigma }} \right\|_1}
\end{equation}
where ${{w}_{{\rm{tr}}}} \in \mathbb{R}_{ +  + }$ is a weighting term, and ${\left\|  \cdot  \right\|_1}$ denotes one-norm. The linearization-based method also suffers from the artificial infeasibility issue \cite{szmuk2020}. The multiple constraints, e.g. state constraints, linearized constraints, and the trust-region constraint, may cause infeasibility. For example, if the problem is linearized based on an unrealistic reference trajectory, the linearized constraints and the trust-region constraint cannot be simultaneously satisfied. To solve this issue, a virtual control term ${{\bm{\mu }}_k} \in {\mathbb{R}^{{n_x}}}$ is added into the discrete-time linearized dynamics (\ref{New dynamics}) according to
\begin{equation}
    \label{Dynamics vc}
    {{\bm{x}}_{k + 1}} = {A_k}{{\bm{x}}_k} + {{\hat B}_k}{{\bm{u}}_k} + {B_k}{{\bm{u}}_{k + 1}} + {\bm{s}_k}{t_f} + {\bm{c}_k}+{{\bm{\mu }}_k}
\end{equation}
The virtual control should be only used for constraint satisfaction. Thus, the cost function is augmented with a large weighting term $w_{{\rm{vc}}}\in {\mathbb{R}_{++}}$ as
\begin{equation}
    \label{Jvc}
    {J_{{\rm{vc}}}} = {w_{{\rm{vc}}}}{\left\| V \right\|_1}
\end{equation}
where $V \buildrel \Delta \over = {\left[ {\begin{array}{*{20}{c}}
{{{\bm{\mu }}_1}}&{{{\bm{\mu }}_2}}& \ldots &{{{\bm{\mu }}_{N - 1}}}
\end{array}} \right]} \in {\mathbb{R}^{{n_x} \times (N - 1)}}$. 

The convex optimization subproblem is summarized as Problem 2.

\begin{table*}[!ht]
\centering 

\begin{tabular}{p{2.5cm}<{\raggedright}cccc} \toprule[1pt] 
\multicolumn{5}{p{15cm}<{\raggedright}}{\emph{Problem 2}: Find the optimization variables ${{t_f},\left\{ {{{\bm{x}}_k}} \right\}_{k = 1}^N,\left\{ {{{\bm{u}}_k}} \right\}_{k = 1}^N,V,{\bm{\sigma }}}$ to solve the convex optimal control subproblem.}  \\ \hline
\emph{Cost function} &\multicolumn{2}{p{8cm}<{\raggedright}}{$\mathop {\min }\limits_{{t_f},\left\{ {{{\bm{x}}_k}} \right\}_{k = 1}^N,\left\{ {{{\bm{u}}_k}} \right\}_{k = 1}^N,V,{\bm{\sigma }}} \quad  - m\left( {{\tau_N}} \right)+{J_{{\rm{tr}}}(\bm{\sigma })}+{J_{{\rm{vc}}}(V)}$} & & See Eqs. (\ref{Jtr}) and (\ref{Jvc})   \\ \hline

\multicolumn{1}{l}{\multirow{3}{*}{\emph{Boundary conditions}}}&\multicolumn{2}{p{7cm}<{\raggedright}}{}&&\\
&	\multicolumn{2}{p{8cm}<{\raggedright}}{$m\left( {{\tau_0}} \right) = {m_0},{{\bm{r}}_{{\cal I}}}\left( {{\tau_0}} \right) = {{\bm{r}}_{{{\cal I}}{\rm{0}}}}, {{\bm{v}}_{{\cal I}}}\left( {{\tau_0}} \right) = {{\bm{v}}_{{{\cal I}}{\rm{0}}}}$ ${{\bm{q}}_{{{\cal I}{\cal B}}}}\left( {{\tau_0}} \right) = {{\bm{q}}_0},{{\bm{\omega }}_{{\cal B}}}\left( {{\tau_0}} \right) = {{\bm{\omega }}_{{{\cal B}}{\rm{0}}}},{{\bm{T}}_{\cal{B}}}\left( {{\tau_0}} \right) ={{\bm{T}}_{{\cal{B}}\rm{0}}}$} &	 & See Eqs. (\ref{T0}-\ref{eq17}) \\ 
& \multicolumn{2}{p{8cm}<{\raggedright}}{${{\bm{r}}_{{\cal I}}}\left( {{\tau_N}} \right) = \bm{0}, {{\bm{v}}_{{\cal I}}}\left( {{\tau_N}} \right) = \bm{0},{{\bm{q}}_{{{\cal I}{\cal B}}}}\left( {{\tau_N}} \right) = {{\bm{q}}_i},{{\bm{\omega }}_{{\cal B}}}\left( {{\tau_N}} \right) = \bm{0}$} & & \\ \hline

\multicolumn{1}{l}{\multirow{3}{*}{\emph{Dynamics}}}&\multicolumn{2}{p{7cm}<{\raggedright}}{}&&\\
&	\multicolumn{2}{p{7cm}<{\raggedright}}{${{\bm{x}}_{k + 1}} = {A_k}{{\bm{x}}_k} + {{\hat B}_k}{{\bm{u}}_k} + {B_k}{{\bm{u}}_{k + 1}} + {\bm{s}_k}{t_f} + {\bm{c}_k}+{{\bm{\mu }}_k}$} & $\forall k \in \{ 1,2, \ldots ,N - 1\} $	 & See Eq. (\ref{Dynamics vc}) \\ \hline

\multicolumn{1}{l}{\multirow{3}{*}{\emph{State constraints}}}&\multicolumn{2}{p{7cm}<{\raggedright}}{}&&\\
&	\multicolumn{2}{p{7cm}<{\raggedright}}{${m_{{\rm{min}}}} \le m(\tau_k)$} &	 & See Inequality (\ref{eq7}) \\   
& \multicolumn{2}{p{7cm}<{\raggedright}}{${\left\| {{H_c}{{\bm{r}}_{{\cal I}}}(\tau_k)} \right\|} \le \cot {\gamma _c}{\bm{e}} \cdot {{\bm{r}}_{{\cal I}}}(\tau_k)$} &	 & See Inequality (\ref{eq8}) \\ 
&\multicolumn{2}{p{7cm}<{\raggedright}}{$2{\left\| {{H_t}{{\bm{q}}_{{{\cal I}{\cal B}}}}(\tau_k)} \right\|^2} \le 1 - \cos {\theta _{\max }}$} & $\forall k \in \{ 1,2, \ldots ,N\} $	 & See Inequality (\ref{eq10})\\ 
&\multicolumn{2}{p{7cm}<{\raggedright}}{${\left\| {{{\bm{\omega }}_{{\cal B}}}(\tau_k)} \right\|_\infty} \le {\omega _{\max }}$} &	 & See Inequality (\ref{eq12}) \\ \hline

\multicolumn{1}{l}{\multirow{3}{*}{\emph{Control constraints}}}&\multicolumn{2}{p{7cm}<{\raggedright}}{}&&\\
&	\multicolumn{2}{p{7cm}<{\raggedright}}{${\left\|  {H_c}{{{\bm{T}}_{{\cal B}}}(\tau_k)} \right\|} \le \tan {\vartheta _{\max }}{\bm{e}} \cdot {{\bm{T}}_{{\cal B}}}(\tau_k)$} &	 & See Inequality (\ref{eq13}) \\   
&\multicolumn{2}{p{7cm}<{\raggedright}}{$\left\|\boldsymbol{T}_{\mathcal{B}}(\tau_k)\right\| \leq T_{\max }$} & $\forall k \in \{ 1,2, \ldots ,N\} $	 & See Inequality (\ref{eq14}) \\
&\multicolumn{2}{p{7cm}<{\raggedright}}{$T_{\min } \leq {H_{\tilde u}}(\tau_k){\bm{u}}(\tau_k )$} &	 & See Eq. (\ref{Linear lower bound}) \\
\bottomrule[1pt] 

\end{tabular}
\end{table*}

\subsection{Sequential Convex Programming Algorithm} \label{SCP algorithm}
As mentioned in Sec. \ref{Convex subproblem}, the SCP algorithm is an iterative algorithm. In each iteration, the algorithm uses a solver for convex optimization to efficiently solve the subproblem Problem 2 \cite{benedikter2021,szmuk2020,reynolds2020,szmuk2018,shen2022}. In this subsection, we will introduce how the algorithm is initialized and terminated.

\emph{1. \quad Initialization}

The initialization approach generates an initial reference trajectory to start the iterative process. The trajectory generated by the initialization approach is only used in the first iteration, and the solution of the previous iteration is used as the reference trajectory in the next iteration to iteratively formulate the convex optimization subproblem.

There are already some proposed initialization approaches, e.g., the straight-line initialization and the 3-DoF initialization \cite{szmuk2020,reynolds2020}. Although the SCP algorithm can handle a wide range of initialization guesses, the quality of guesses greatly affects the convergence performance. Poor guesses may lead to more iterations, which can increase computation time. In this paper, an initial trajectory generator is proposed in Sec. \ref{Initial Trajectory Generator}. 
The proposed method is based on DNN and uses the pre-computed results to train the neural network so that it can give a satisfactory initial trajectory guess. By giving a good initial trajectory guess, the convergence of the iterative algorithm can be accelerated to reduce the computation time. The details of the proposed initial trajectory generator will be introduced in Sec. \ref{Initial Trajectory Generator}. 

\emph{2. \quad Nondimensionalization}

A numerical solver may encounter issues associated with machine precision and sensitivity due to the different magnitudes of variables in the optimization problem. A nondimensionalization is performed by scaling the optimization variables via
\begin{subequations}
\label{Nondimentionalization}
\begin{gather}
    \bar m = m/{U_M},{{{\bm{\bar r}}}_{{\cal I}}} = {{\bm{r}}_{{\cal I}}}/{U_L},{{{\bm{\bar v}}}_{{\cal I}}} = {{\bm{v}}_{{\cal I}}}({U_T}/{U_L})\\
    {{{\bm{\bar T}}}_{{\cal B}}} = {{\bm{T}}_{{\cal B}}}\frac{{U_T^2}}{{{U_M}{U_L}}}
\end{gather}
\end{subequations}
where ${U_M} \buildrel \Delta \over = {m_0}$, ${U_L} \buildrel \Delta \over = \left\| {{{\bm{r}}_{{{\cal I}}0}}} \right\|$, and ${U_T} \buildrel \Delta \over = 1\,{\rm{s}}$ are the scaling units.

\emph{3. \quad Convergence Criteria}

The iteration process terminates when convergence criteria are met. The convergence criteria are given as
\begin{subequations}
\label{Criteria1}
\begin{align}
    {J_{{\rm{tr}}}}(\bm{\sigma} ) & \le {\epsilon _{{\rm{tr}}}} \label{Jtr tolerance} \\
    {J_{{\rm{vc}}}}(V) & \le {\epsilon _{{\rm{vc}}}} \label{Jvc tolerance}
\end{align}
\end{subequations}
where ${\epsilon _{{\rm{tr}}}} \in {\mathbb{R}_{ +  + }}$ and ${\epsilon _{{\rm{vc}}}} \in {\mathbb{R}_{ +  + }}$ are the convergence tolerances, which can be user-specified. The convergence criterion (\ref{Jtr tolerance}) measures the difference between the solutions of two consecutive iterations. Additionally, the criterion (\ref{Jvc tolerance}) guarantees that the solution meets the dynamics. 

Comparing the maximum difference in the solutions between two consecutive iterations is another approach. The algorithm terminates when the difference is less than a tolerance. The criterion can be given as
\begin{equation}
    \label{X tolerance}
    \mathop {\max }\limits_{k \in \{ 1,2, \ldots ,N\} } {\left\| {\left( {{{\bm{x}}_k} - {{{\bm{\tilde x}}}_k}} \right)} \right\|_\infty } < {\epsilon _x}
\end{equation}
where ${\epsilon _x} \in {\mathbb{R}_{ +  + }}$ is the tolerance. Using different numerical solvers, even solving the same problem, can result in different values for criteria (\ref{Jtr tolerance}-\ref{Jvc tolerance}) \cite{reynolds2020}. The criterion (\ref{X tolerance}) is uniformly functional across different solvers. Criteria (\ref{Criteria1}) are more conducive to obtaining the dynamically feasible solution, but require more iterations to converge, while criterion (\ref{X tolerance}) is more suitable for online application. In Sec. \ref{Results}, the criteria (\ref{Criteria1}) are used to construct the data set, and the criterion (\ref{X tolerance}) is used to test guidance performance.

In this paper, we combine SCP algorithms with DNN to improve the performance of SCP algorithms, instead of using neural networks as controllers brutally. The proposed algorithm is detailed as Algorithm \ref{alg:alg SCP} in Sec. \ref{Initial Trajectory Generator} together with the proposed initial trajectory generator.

\section{Initial Trajectory Generator} \label{Initial Trajectory Generator}
\begin{figure*}[!ht]
\centering
\includegraphics[width=18cm]{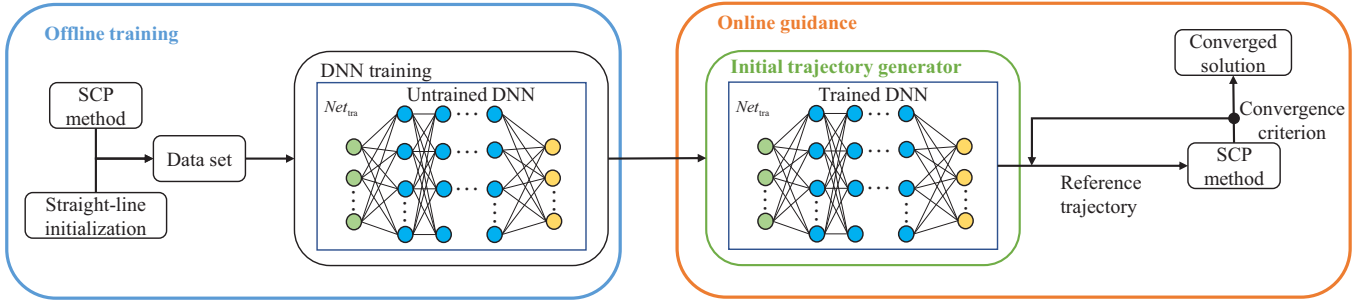}
\caption{Overview of the proposed scheme.}
\label{overview}
\end{figure*}

In this section, the initial trajectory generator based on the DNN is proposed, which can significantly reduce the computation time of the SCP algorithm. In general, we use the straight-line initialization to initialize the SCP algorithm. The algorithm is used to solve guidance problems with various initial conditions. The obtained solutions are collected to construct a data set. The data set is used to train the DNN-based generator, so that the generator can give a satisfactory initial guess trajectory. The overview of the proposed scheme is shown in Fig. \ref{overview}. In Sec. \ref{Data set}, the construction of the data set is introduced. The structure of the DNN-based generator is presented in Sec. \ref{Generator structure}. In Sec. \ref{Generator training}, the details on how to train the DNN-based generator are given.

\subsection{Data Set} \label{Data set}

\emph{1. \quad Straight-line Initialization}

The data set is constructed by solving various guidance problems. Thus, a baseline initialization approach is introduced. The straight-line initialization method is presented as
\begin{subequations}
\begin{gather}
\tilde{\boldsymbol{x}}_k=\left(\frac{N-k}{N-1}\right) \tilde{\boldsymbol{x}}_{\mathrm{ini}}+\left(\frac{k-1}{N-1}\right) \tilde{\boldsymbol{x}}_{\mathrm{fin}} \\
{\tilde {\bm{u}} _k} = \frac{{{T_{\max }} - {T_{\min }}}}{2}{\bm{e}}
\end{gather}
\end{subequations}
where ${\widetilde {\bm{x}}_{{\rm{ini}}}} \buildrel \Delta \over = {\left[ {\begin{array}{*{20}{l}}
{{m_0}}&{{\bm{r}}_{{{\cal I}}0}^T}&{{\bm{v}}_{{{\cal I}}0}^T}&{{\bm{q}}_{i}^T}&{{\bm{\omega }}_{{{\cal B}}{\rm{0}}}^T}
\end{array}} \right]^T}$, ${\widetilde {\bm{x}}_{{\rm{fin}}}} \buildrel \Delta \over = {\left[ {\begin{array}{*{20}{l}}
{{m_{{\rm{min }}}}}&{{{\bm{0}}^T}}&{{{\bm{0}}^T}}&{{\bm{q}}_{i}^T}&{{{\bm{0}}^T}}
\end{array}} \right]^T}$, and $k \in \{ 1,2, \ldots ,N\} $.

\emph{2. \quad Data Set Construction}

Disturbing the initial state of the vehicle with the uniformly distributed stochastic parameter ${\bm{\xi }} \in {\mathbb{R}^{{n_x}}}$, the various guidance problems with various initial states can be obtained. The disturbed initial state can be expressed as
\begin{equation}
    \label{Disturb}
    {\bm{x}}_1^n = {{\bm{x}}_1} + {{\bm{\xi }}^n},\quad \forall n \in \{ 1,2, \ldots ,{N_{{\rm{tra}}}}\} 
\end{equation}
where the superscript $n$ denotes the $n$-th perturbation, ${N_{\rm{tra}}}$ is the number of trajectories in the data set, and ${{\bm{x}}_1} \buildrel \Delta \over = {\left[ {\begin{array}{*{20}{l}}
{{m_0}}&{{\bm{r}}_{{{\cal I}}0}^T}&{{\bm{v}}_{{{\cal I}}0}^T}&{{\bm{q}}_{0}^T}&{{\bm{\omega }}_{{{\cal B}}{\rm{0}}}^T}
\end{array}} \right]^T}$. Each perturbation generates a guidance problem. The SCP algorithm presented in Sec. \ref{Convex programming} is used to solve the various guidance problems. As a result, the data set contains ${N_{\rm{tra}}}$ optimized trajectories. The constructed data set is shown in Sec. \ref{Results}.

\subsection{Trajectory Generator Structure} \label{Generator structure}

\begin{figure*}[!ht]
\centering
\includegraphics[width=18cm]{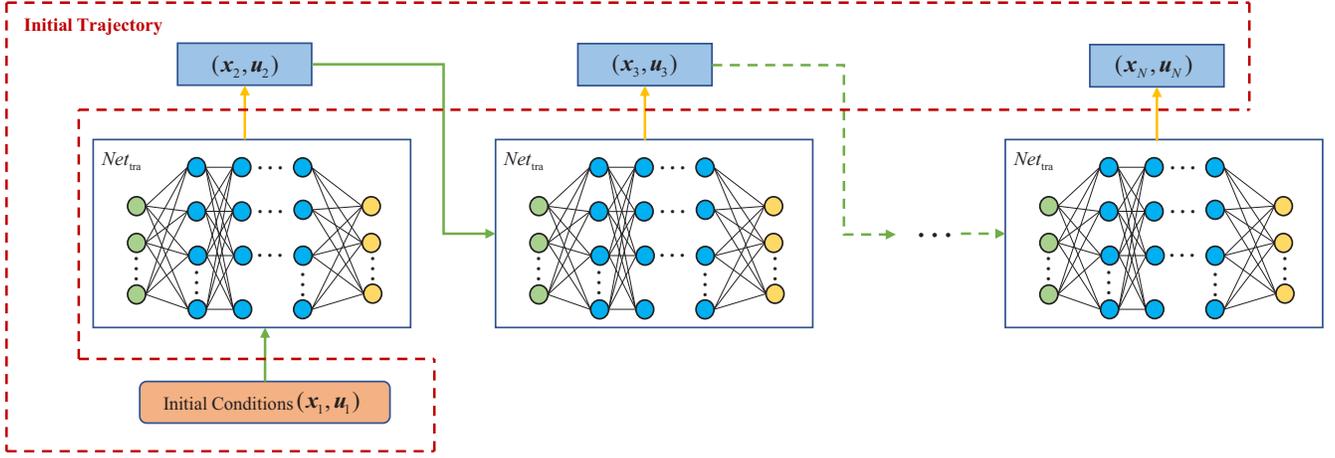}
\caption{Illustration of the initial trajectory generator.}
\label{fig1}
\end{figure*}

Motivated by the sequence model prediction in natural language processing \cite{zhang2021dive,goodfellow2016deep}, the initial trajectory is modeled as a sequence model as shown in Fig. \ref{fig1}. A frame of the trajectory is defined as $({{\bm{x}}_k},{{\bm{u}}_k})$, for $k \in \{ 1,2, \ldots ,N\} $. The DNN is used to recurrently predict each frame of the trajectory until an N-frame trajectory is generated. Given the last frame, the DNN can predict next frame of the trajectory according to

\begin{equation}
    \label{DNN}
    ({{\bm{x}}_{k + 1}},{{\bm{u}}_{k + 1}}) = Net_{{{\rm{tra}}}}({{\bm{x}}_k},{{\bm{u}}_k}),\ \forall k \in \{ 1,2, \ldots ,N - 1\} 
\end{equation}
The generated trajectory can be used as the initial trajectory of SCP. In Sec. \ref{Results}, one can see that this initialization method can significantly reduce the computation time. The SCP algorithm together with the initial trajectory generator is summarized in Algorithm \ref{alg:alg SCP}.

\begin{algorithm}[!ht]
\caption{SCP with Initial Trajectory Generator.}\label{alg:alg SCP}
\begin{algorithmic}
\STATE 
\STATE {\textsc{INITIAL TRAJECTORY GENERATOR}}$(\tilde{\bm{x}}_1,\tilde{\bm{u}}_1)$
\STATE \hspace{0.5cm}\textbf{for} $k \in \{ 1,2, \ldots ,N - 1\}   $ \textbf{do}
\STATE \hspace{1cm}$ \textbf{compute } ({{\tilde{\bm{ x}}}_{k + 1}},{{\tilde{\bm{ u}}}_{k + 1}}) = Ne{t_{{\rm{tra}}}}({{\tilde{\bm{ x}}}_k},{{\tilde{\bm{ u}}}_k})$
\STATE \hspace{0.5cm}\textbf{end for}
\STATE \hspace{0.5cm}\textbf{give initial guess} ${{{\tilde t}_f}}$
\STATE \hspace{0.5cm}\textbf{return} $({{{\tilde t}_f}},\tilde{\bm{x}},\tilde{\bm{u}})$
\STATE 
\STATE {\textsc{SEQUENTIAL CONVEX PROGRAMMING}}$({{{\tilde t}_f}},\tilde{\bm{x}},\tilde{\bm{u}})$
\STATE \hspace{0.5cm}\textbf{while} \emph{not converged} \textbf{do}
\STATE \hspace{1cm}\textbf{use} $({{{\tilde t}_f}},\tilde{\bm{x}},\tilde{\bm{u}})$ to construct Problem 2
\STATE \hspace{1cm}\textbf{solve} Problem 2 to get $(t_f,\bm{x},\bm{u},V,\bm{\sigma})$
\STATE\hspace{1cm}\textbf{if} criterion is satisfied according to (\ref{Jtr tolerance}-\ref{X tolerance})
\STATE\hspace{1.5cm}\emph{converged}
\STATE\hspace{1cm}\textbf{end if}
\STATE \hspace{1cm}$ ({{{\tilde t}_f}},\tilde{\bm{x}},\tilde{\bm{u}}) \gets  (t_f,\bm{x},\bm{u}) $
\STATE \hspace{0.5cm}\textbf{end while}
\STATE \hspace{0.5cm}\textbf{return}  $(t_f,\bm{x},\bm{u})$
\end{algorithmic}
\label{alg1}
\end{algorithm}

It is important to remark that since the initial trajectory generator is a sequence model predictor, it is independent of the dynamics of the vehicle itself, i.e., the generator only recurrently predicts the state and control on each time stamp. Thus, the generator can be applied to other problems by only adjusting the input and output dimensions of the network. In this paper, although we have only studied the powered landing guidance problem, which contains 6-DoF dynamics with nonlinearity, multiple nonlinear constraints, etc., the generator can be applied to any SCP-based guidance method to reduce the computation time.

The structure of the DNN is very important for the performance of the generator. A DNN consists of an input layer, several hidden layers, and an output layer. In this paper, a variety of structures are tested in Sec. \ref{Results}. According to the test results, a five-hidden-layers structure of 256 units per layer is finally selected. Further, the Rectified Linear Unit (ReLU) is adopted as the activation function, which can be expressed as
\begin{equation}\label{ReLU}
{\kappa_j}(x) = \max \left( {x,0} \right),\quad \forall j \in \{1,2, \ldots ,{{N_L}-1}\}
\end{equation}
where $\kappa_j$ means the activation function of the $j$-th layer, and the DNN has $N_L$ layers in total. 

In the original input vector, there is a wide variety of data types, e.g., mass, velocities, and angle rates. Since there is no prior knowledge about which features will be more relevant, it should be avoided that the weights assigned to some features are greater than that of other features. Thus, data normalization is required to achieve a satisfactory result in the training process \cite{shi2021,zhang2021dive}. The input data $\bm{x}$ is normalized as
\begin{equation}\label{Normalization}
\bm{x} \leftarrow \frac{{\bm{x} - E(\bm{x})}}{{De(\bm{x})}}
\end{equation}
where $E(\bm{x})$ and $De(\bm{x})$ denote the mean and standard deviation of $\bm{x}$. The input features are rescaled to zero mean and unit variance. The input of the DNN is a normalized trajectory frame, and the output is the normalized trajectory frame in the next discrete point. Hence, the input and output of the DNN are both a $17\times1$ vector.

\subsection{Trajectory Generator Training} \label{Generator training}

In the training process, the mean squared error (MSE) is utilized as the loss function, which can be expressed as
\begin{equation}\label{Loss}
L = \frac{1}{{{N_b}}}\sum\limits_{i = 1}^{{N_b}} ({ {{\hat y}_i - {y^ * }} })^2
\end{equation}
where $N_b$ is the number of training samples, $\hat y$ is the final output of the DNN, and $y^ *$ is the corresponding target output. Further, the Adam method is employed as the optimizer \cite{kingma2014adam}. Feeding the DNN with the training data set, the optimizer can update the DNN in each epoch to reduce the loss function.

The weight decay technique is also introduced in the training process. The weight decay technique is a regularization technique, which mitigates overfitting by adding a penalty term to the loss function (\ref{Loss}) \cite{zhang2021dive}. The loss function is now given by
\begin{equation}\label{NewLoss}
{L_{{\rm{new}}}} = L + \frac{\lambda }{2}{\left\| {\bm{w}}_{\rm{net}} \right\|^2}
\end{equation}
where ${\bm{w}}_{\rm{net}}$ is the weight vector, and $\lambda$ is the non-negative regularization constant, which characterizes the tradeoff.

\section{Results} \label{Results}
In this section, the test results are presented to demonstrate the effectiveness and performance of the proposed guidance method. In Sec. \ref{Dataset results}, the constructed data set is presented. In Sec. \ref{Training results}, the details of the training process are introduced. In Sec. \ref{Guidance performance}, the performance of the proposed guidance method is illustrated via simulations. The Monte Carlo analysis is performed to test the performance of the algorithm. The proposed algorithm is also compared with the state-of-the-art SCP algorithm to highlight the improvement. The test results are carried out on a desktop computer with an Intel Core i7-12700F processor. The test results are obtained using Python implementation. The CVXPY and MOSEK solver are used to solve the convex optimization problem \cite{diamond2016cvxpy,agrawal2018rewriting,aps2020mosek}.

\subsection{Data Set}\label{Dataset results}
\begin{figure}[!t]
\centering
\includegraphics[width=8cm]{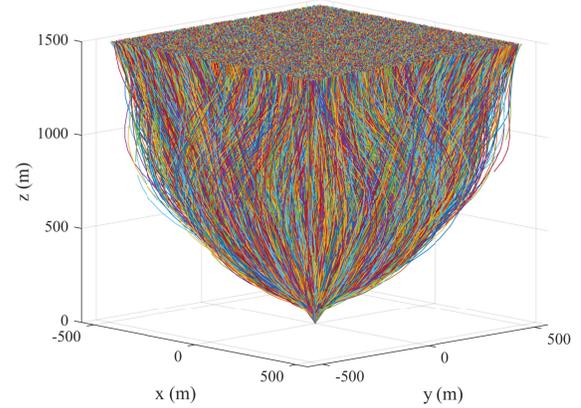}
\caption{Data set.}
\label{fig2}
\end{figure}
The data set is constructed via the SCP method mentioned in Sec. \ref{Convex programming} and \ref{Data set}. The weight parameters in Eqs. (\ref{Jtr}) and (\ref{Jvc}) are selected as ${{w}_{{\rm{tr}}}}=0.5$ and ${{w}_{{\rm{vc}}}}=1 \times {10^5}$, respectively.  The convergence tolerances $\epsilon_{\rm{tr}}$ and $\epsilon_{\rm{vc}}$ in Eqs. (\ref{Jtr tolerance}) and (\ref{Jvc tolerance}) are both selected as $5 \times 10^{-4}$. The initial conditions and parameters of the landing problem are listed in Table \ref{Initial conditions}. The vehicle's moment of inertia is diag($[4 \times 10^6, 4 \times 10^6, 1 \times 10^5]$) $\text{kg} \cdot \text{m}^2$.
\begin{table}[!hb]
  \centering
  \caption{Initial Conditions and Parameters}
  \label{Initial conditions}
  \begin{tabular}{|c|c|c|c|}
    \hline
    \bf{Parameter} &  \bf{Value} & \bf{Parameter} &  \bf{Value}\\ \hline
    ${m_0}$ (kg) &  $30000$ & ${T _{\max }}$ (N)  & $800000$\\ \hline
    ${{\bm{r}}_{{{\cal I}}{\rm{0}}}}$ (m) &  ${\left[ 0 \ 0\ 1500 \right]^T}$ & ${T _{\min }}$ (N)  & $320000$\\ \hline
    ${{\bm{v}}_{{{\cal I}}{\rm{0}}}}$ (m/s) & ${\left[ 0 \ 0\ -80 \right]^T}$& ${{{\tilde t}_f}}$ (s)  & $18$\\ \hline
    ${{\bm{q}}_0}$ & ${\left[ 0 \ 0\ 0\ 1 \right]^T}$ & ${\bm{g}_\mathcal{I}}$ ($\text{m} / \text{s}^2$) & ${\left[ 0 \ 0\ -9.81 \right]^T}$\\ \hline
    ${{\bm{\omega }}_{{{\cal B}}{\rm{0}}}}$ (deg/s) & ${\left[ 0 \ 0\ 0 \right]^T}$ & $S_A$ ($\text{m}^2$)  & $10$\\ \hline
    ${{\bm{T}}_{{\cal{B}}\rm{0}}}$ (N)  & ${\left[ 0 \ 0\ {{T_{\min }}} \right]^T}$& $C_A$   & diag($[3, 3, 1]$)\\ \hline
    ${m_{{\rm{min}}}}$ (kg)  & $22000$& $\rho$ ($\text{kg} / \text{m}^3$)  & $1.225$\\ \hline
    $N$  & 30 & ${I_{{\rm{sp}}}}$ (s)  & $282$\\ \hline
    ${\omega _{\max }}$ (deg/s)  & $30$& ${P_{{\rm{atm}}}}$ ($\text{Pa}$)  & $0$\\ \hline
    ${\gamma _{c }}$ (deg)  & $20$& ${S_{{\rm{ne}}}}$ ($\text{m}^2$)  & $0$\\ \hline
    ${\theta _{\max }}$ (deg)  & $80$& $\bm{d}_{T, \cal{B}}$ (m)  & ${\left[ 0 \ 0\ -14 \right]^T}$\\ \hline
    ${\vartheta _{\max }}$ (deg)  & $20$& $\bm{d}_{A, \cal{B}}$ (m)  & ${\left[ 0 \ 0\ 2 \right]^T}$\\ \hline
    
  \end{tabular}
\end{table}

As discussed in Sec. \ref{Data set}, a uniformly distributed term ${\bm{\xi }} \buildrel \Delta \over = {\left[ {\begin{array}{*{20}{c}}
{{{\bm{\xi }}_m}^T}&{{{\bm{\xi }}_r}^T}&{{{\bm{\xi }}_v}^T}&{{{\bm{\xi }}_q}^T}&{{{\bm{\xi }}_\omega }^T}
\end{array}} \right]^T} \in {\mathbb{R}^{{n_x}}}$ is added to the initial state of the vehicle to generate various guidance problems. The ranges of the random parameters are listed in Table \ref{Random}.
\begin{table}[!htb]
  \centering
  \caption{Ranges of Random Parameters}
  \label{Random}
  \begin{threeparttable} 
  \begin{tabular}{|c|c|}
    \hline
    \bf{Parameter} &  \bf{Range} \\ \hline
    ${{{\bm{\xi }}_m}}$ (kg) &  $0$ \\ \hline
    ${{{\bm{\xi }}_r}}$ (m) &  ${\left[ \left[ -500,500 \right] \ \left[ -500,500 \right]\ 0 \right]^T}$ \\ \hline
    ${{{\bm{\xi }}_v}}$ (m/s) & ${\left[ \left[ -40,40 \right] \ \left[ -40,40 \right]\ \left[ -20,20 \right] \right]^T}$\\ \hline
    ${{{\bm{\xi }}_q}}$ & $Euler2Quater^{*}({\left[ \left[ -30,30 \right] \ \left[ -30,30 \right]\ 0 \right]^T})-\bm{q}_i$ \\ \hline
    ${{{\bm{\xi }}_\omega }}$ (deg/s) & ${\left[ \left[ -20,20 \right] \ \left[ -20,20 \right]\ 0 \right]^T}$\\ \hline
    
  \end{tabular}
  \begin{tablenotes} 
        \footnotesize 
        \item[*] Function that convert Euler angles (deg) to quaternions.
  \end{tablenotes} 
  \end{threeparttable} 
\end{table}

\begin{figure*}[hbt]
\centering
\subfloat[]{\includegraphics[width=8cm]{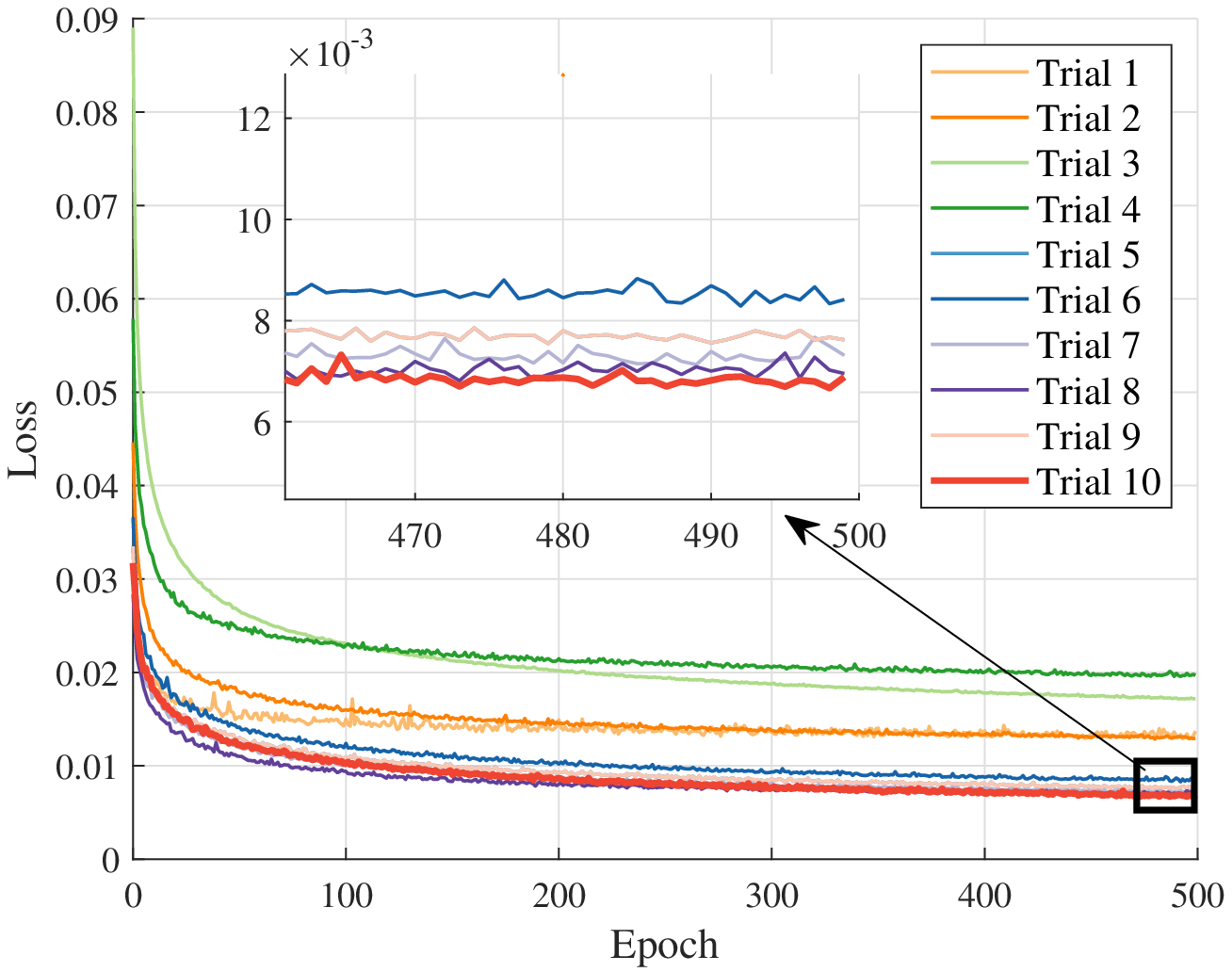}%
\label{fig3a}}
\hfil
\subfloat[]{\includegraphics[width=8cm]{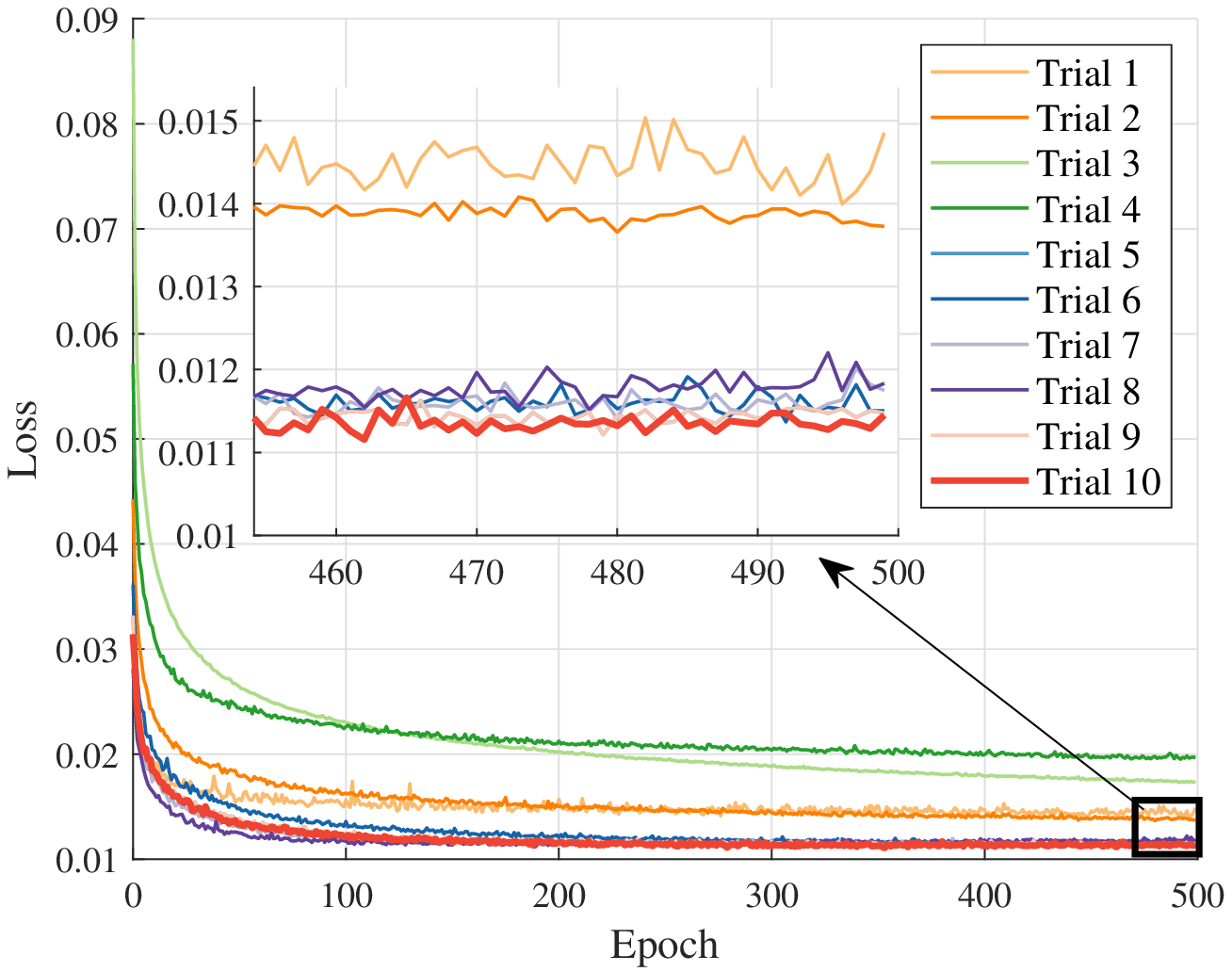}%
\label{fig3b}}
\caption{Training process of trials 1-10. 500 epochs of training are carried out. (a) Training loss. (b) Test loss.}
\label{fig3}
\end{figure*}
\begin{figure*}[!ht]
\centering
\subfloat[]{\includegraphics[width=8cm]{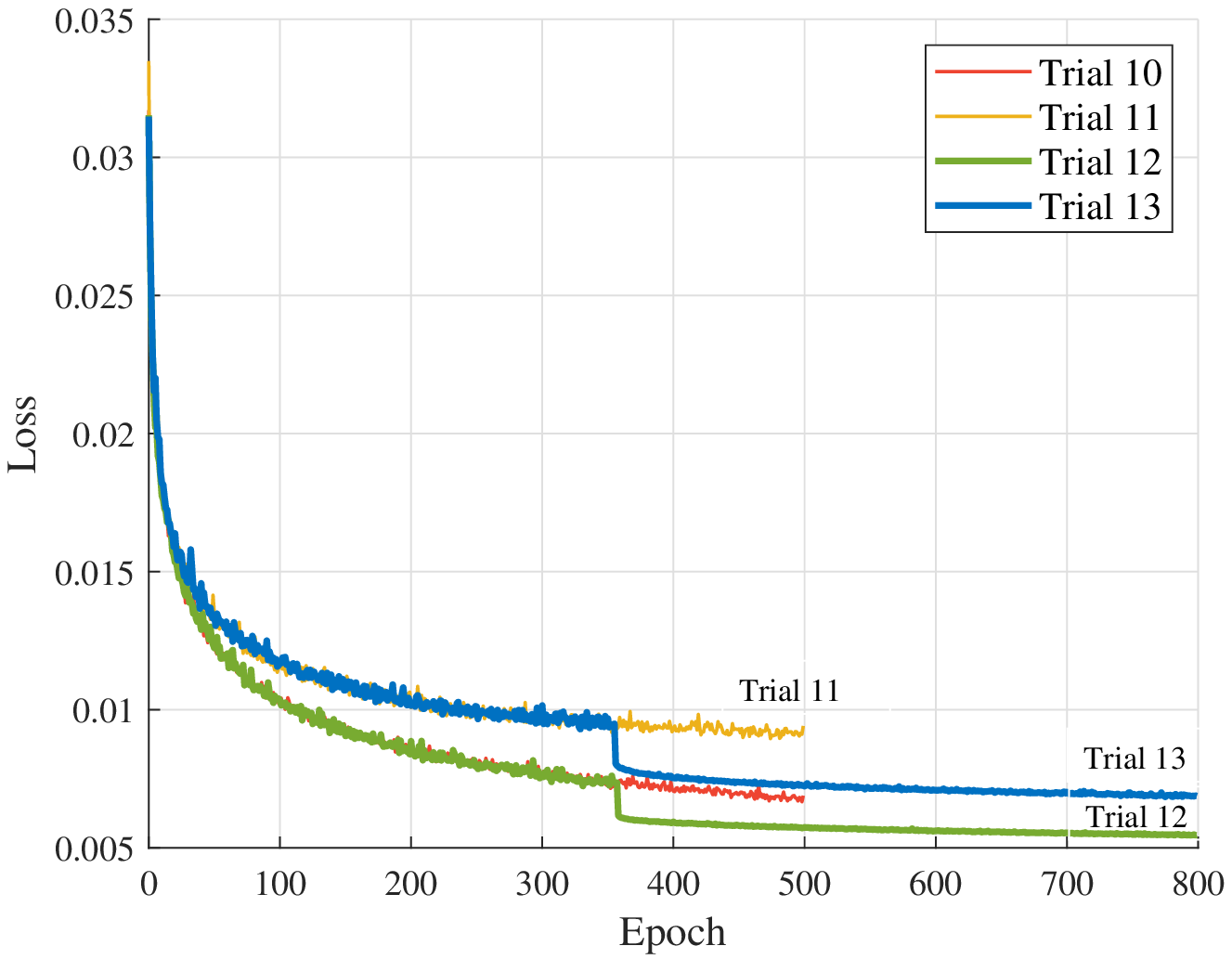}%
\label{fig4a}}
\hfil
\subfloat[]{\includegraphics[width=8cm]{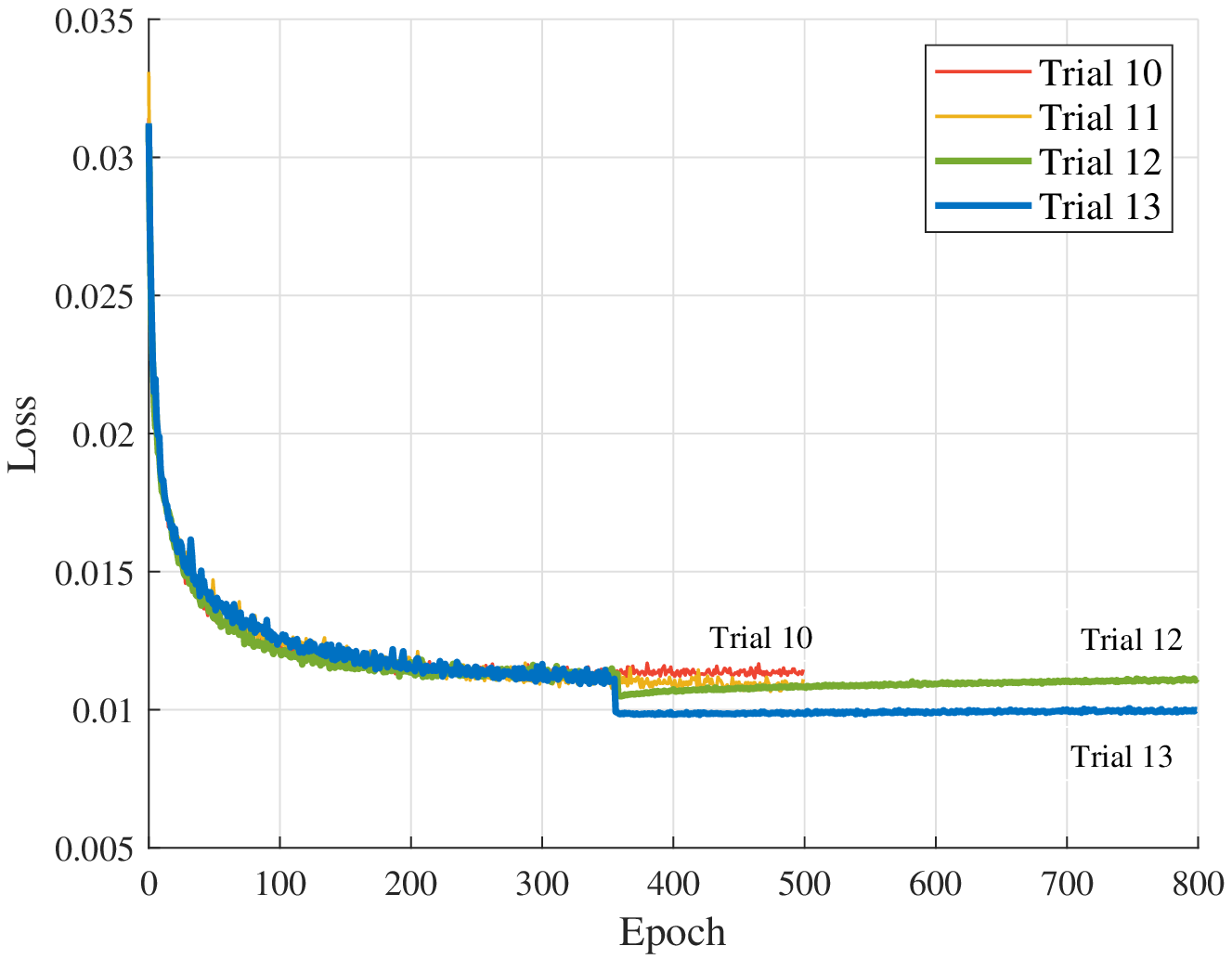}%
\label{fig4b}}
\caption{Training process of trials 10-13. 500 epochs of training are carried out in trials 10 and 11, and 800 epochs of training are carried out in trials 12 and 13. (a) Training loss. (b) Test loss.}
\label{fig4}
\end{figure*}

The data set can be obtained by solving various guidance problems. In this paper, the data set contains 48333 trajectories, 45000 of which are used for training and 3333 for testing. The randomly sampled trajectories with various initial conditions are shown in Fig. \ref{fig2}.

\subsection{Training Results}\label{Training results}

In the training process, to determine the better hyperparameters, the training performance of different hyperparameters is tested. A total of 13 trials are tested, and the information on these trials is listed in Table \ref{Trials}. The loss functions during training are shown in Figs. \ref{fig3} and \ref{fig4}. It should be pointed out that in trials 12 and 13, learning rate decay is used to make the training process converges better. When the training loss is not reduced in 25 epochs of training, the learning rate is reduced ten times. The minimum learning rate is limited to $1 \times 10^{-6}$. In trails 1-11, 500 epochs of training are carried out. But 800 epochs of training are carried out in trials 12 and 13 to test the effect of learning rate decay.
\begin{table}[!htb]
  \centering
  \caption{Information of Trials}
  \label{Trials}
  \begin{threeparttable} 
  \begin{tabular}{|c|c|c|c|c|c|}
    \hline
    \bf{Trial} &  \bf{Learning Rate} &  \bf{${\textbf{Unit}}^{1}$}  &  \bf{{${\textbf{Layer}}^{2}$}}  &  \bf{${\textbf{Batch}}^{3}$}  &  \bf{{${\textbf{WD}}^{4}$}}\\ \hline
    1 & $1 \times 10^{-3}$ & 128 & 5 & 128 & --  \\ \hline
    2 & $1 \times 10^{-4}$ & 128 & 5 & 128 & --    \\ \hline
    3 & $1 \times 10^{-5}$ & 128 & 5 & 128 & --  \\ \hline
    4 & $1 \times 10^{-4}$ & 64 & 5 & 128 & --   \\ \hline
    5 & $1 \times 10^{-4}$ & 256 & 5 & 128 & --  \\ \hline
    6 & $1 \times 10^{-4}$ & 256 & 5 & 256 & --    \\ \hline
    7 & $1 \times 10^{-4}$ & 256 & 5 & 64 & --    \\ \hline
    8 & $1 \times 10^{-4}$ & 256 & 5 & 32 & --  \\ \hline
    9 & $1 \times 10^{-4}$ & 256 & 4 & 128 & --   \\ \hline
    10 & $1 \times 10^{-4}$ & 256 & 6 & 128 & --  \\ \hline
    11 & $1 \times 10^{-4}$ & 256 & 6 & 128 & $1 \times 10^{-5}$    \\ \hline
    12 & $1 \times 10^{-4}$ & 256 & 6 & 128 & --    \\ \hline
    13 & $1 \times 10^{-4}$ & 256 & 6 & 128 & $1 \times 10^{-5}$  \\ \hline
    
  \end{tabular}
  \begin{tablenotes} 
        \footnotesize 
        \item[1] Number of units in each layer of the DNN.
        \item[2] Number of DNN layers.
        \item[3] Batch size.
        \item[4] Constant $\lambda$ in weight decay (see Eq. (\ref{NewLoss})).
  \end{tablenotes} 
  \end{threeparttable} 
\end{table}

As shown in Fig. \ref{fig3}, the training loss and test loss of trail 10 are minimal throughout the test. Further, learning rate decay is used to make the loss converge better. In Fig. \ref{fig4}, although trial 12, which uses the same hyperparameters as trial 10, achieves lower training loss, the test loss of trial 12 is close to that of trial 10. Hence, there is overfitting. Then, the weight decay mentioned in Sec. \ref{Generator training} is used in the training process to alleviate the overfitting phenomenon. Trial 13 achieves the best test loss and acceptable training loss results, which indicates that the DNN may have satisfactory generalization ability. Therefore, in Sec. \ref{Guidance performance}, the DNN obtained from trial 13 will be used as the initial trajectory generator.

\subsection{Guidance Performance}\label{Guidance performance}
\begin{figure*}[!htb]
\centering
\subfloat[]{\includegraphics[width=8cm]{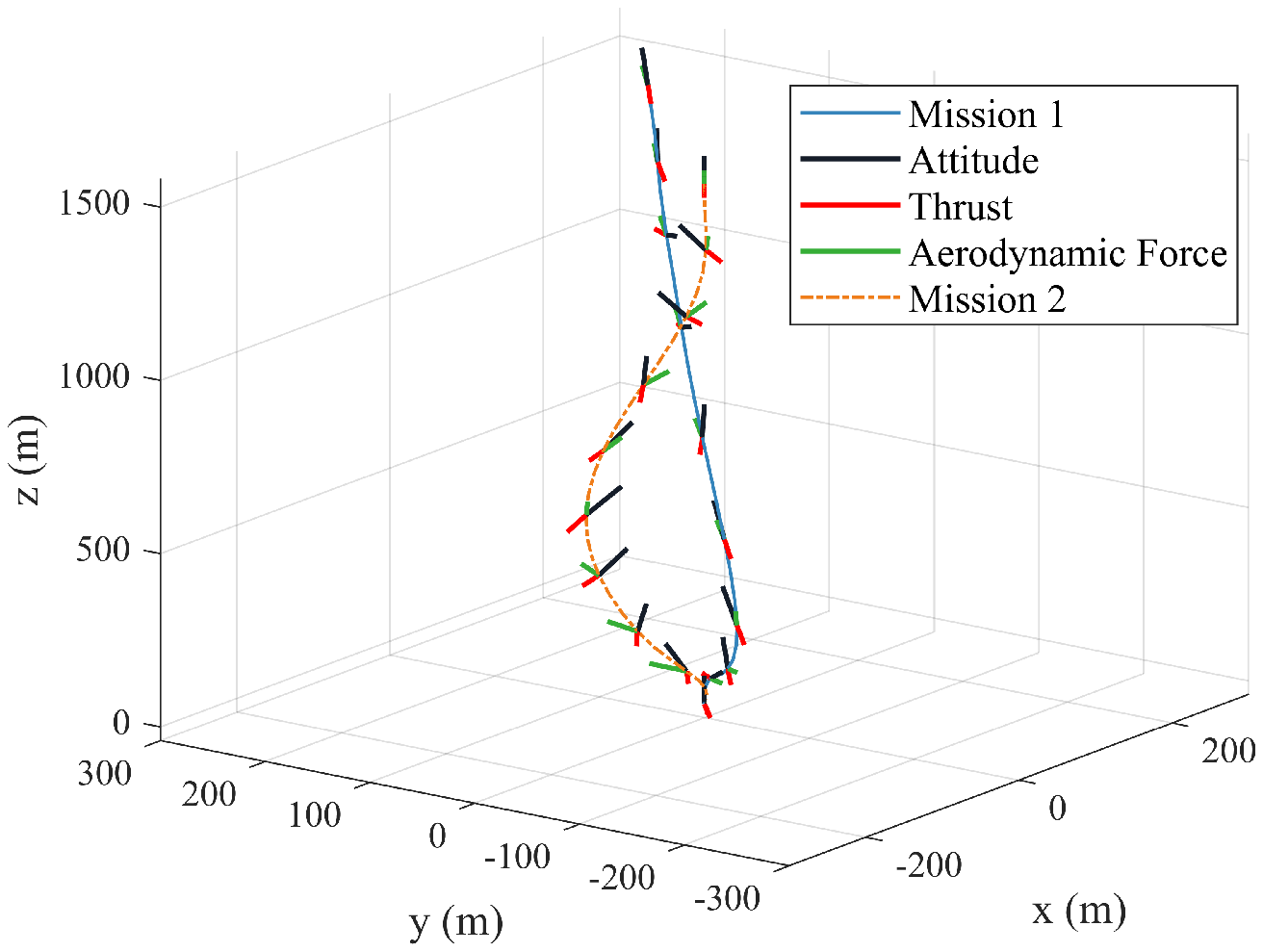}%
\label{fig5a}}
\hfil
\subfloat[]{\includegraphics[width=8cm]{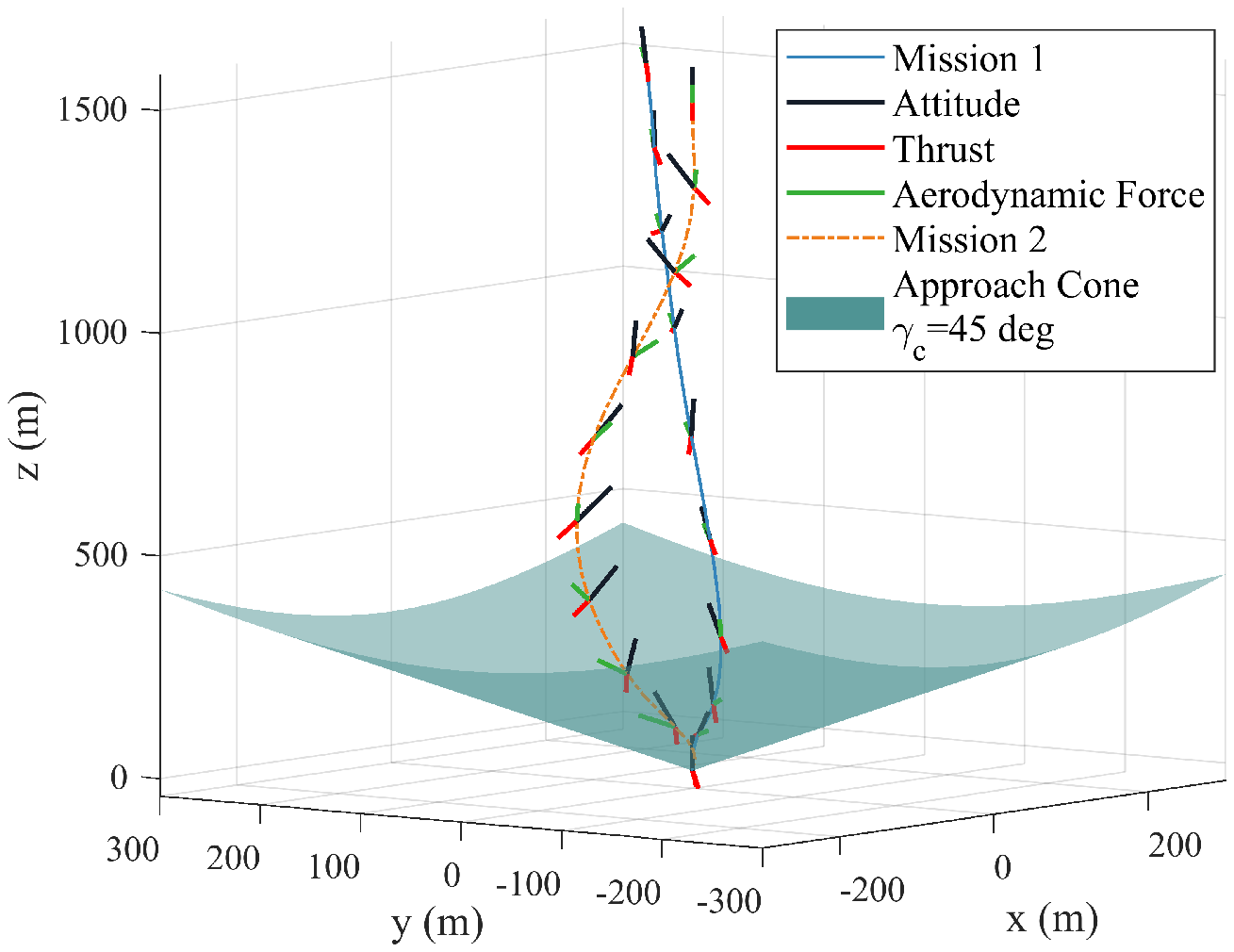}%
\label{fig5b}}
\caption{Solutions of the missions. (a) Approach cone is not displayed. (b) Approach cone is displayed.}
\label{fig5}
\end{figure*}
\begin{figure*}[!ht]
\centering
\subfloat[]{\includegraphics[width=8cm]{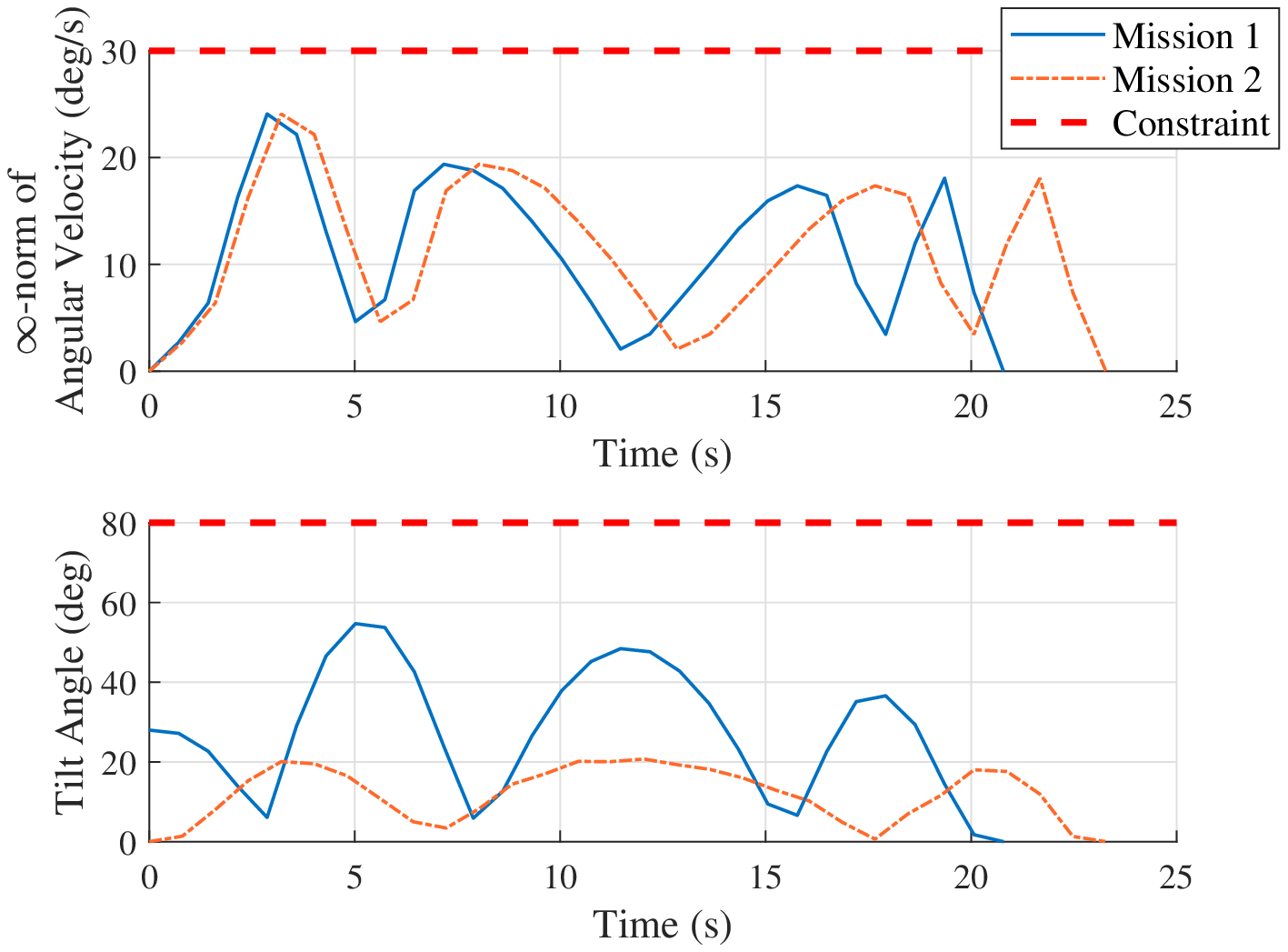}%
\label{fig6a}}
\hfil
\subfloat[]{\includegraphics[width=8cm]{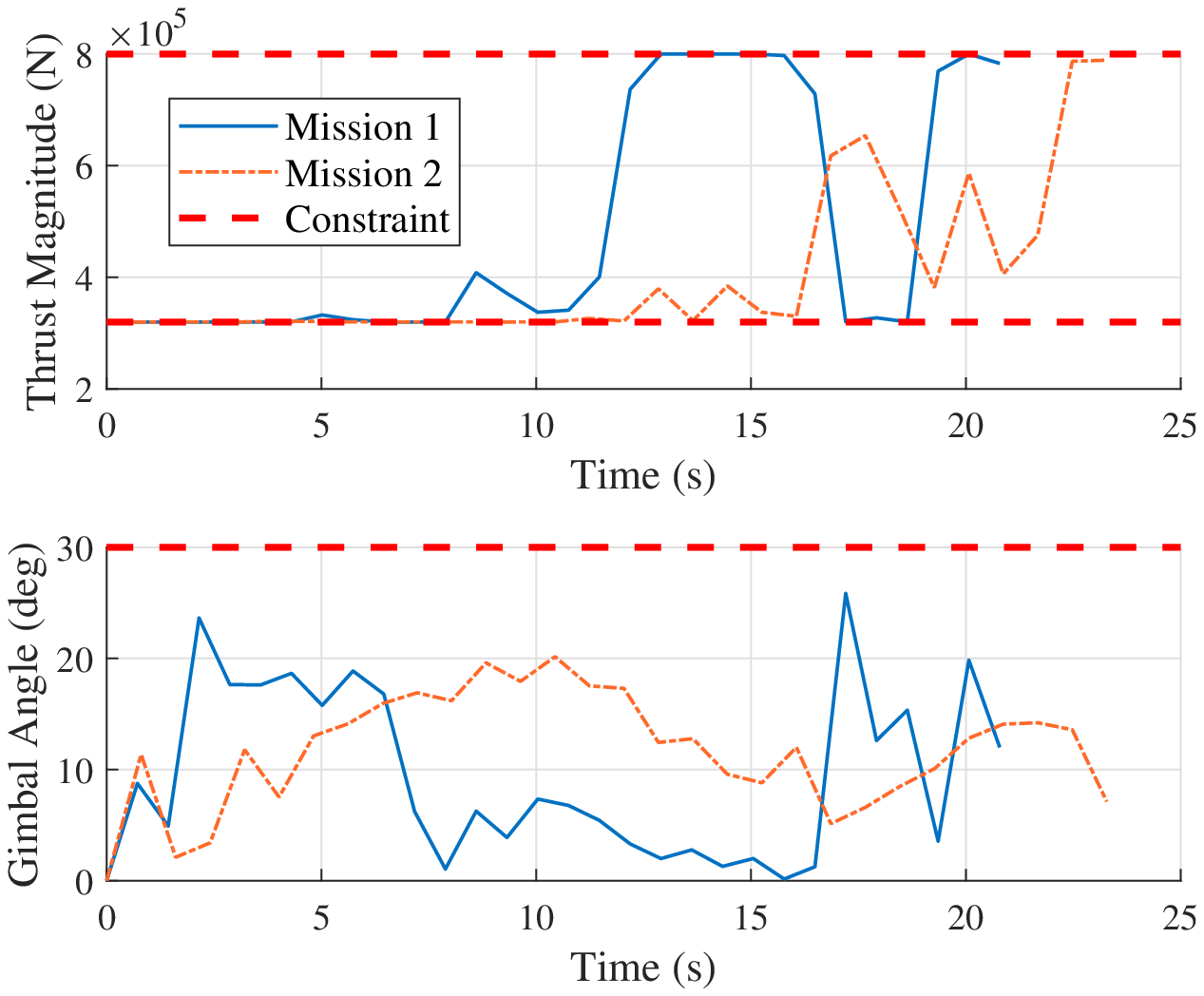}%
\label{fig6b}}
\caption{Constraints in the missions. (a) State constraints. (b) Control constraints.}
\label{fig6}
\end{figure*}
The proposed guidance method is validated by two missions with different initial conditions. The initial conditions and parameters are listed in Table \ref{Two missions}, and the parameters not listed are the same as those in Table \ref{Initial conditions}. The convergence tolerance ${\epsilon _x}$ in Eq. (\ref{X tolerance}) is selected as $1 \times 10^{-2}$. The solutions of these two missions are shown in Fig. \ref{fig5}. As displayed in Fig. \ref{fig5a}, the proposed algorithm can guide the vehicle to the landing site. And the approach cone constraint can be satisfied, which can be observed in Fig. \ref{fig5b}. In addition, the details of the state constraints and control constraints are shown in Fig. \ref{fig6}. One can ensure that these constraints are all satisfied. And the minimum mass constraint can be ensured by observing that the final mass of Mission 1 is 26403.8 kg, and the final mass of Mission 2 is 26697.4 kg. In Fig. \ref{fig6b}, it should also be noted that the obtained solutions exhibit behavior that is close to bang-bang control, while early works noted that the solution to 3-DoF fuel-optimal powered landing guidance also exhibited bang-bang behavior \cite{1105758,marec2012optimal}.

\begin{table}[!htb]
  \centering
  \caption{Initial Conditions of Missions}
  \label{Two missions}
  \begin{threeparttable} 
  \begin{tabular}{|c|c|c|}
    \hline
    \bf{Parameter} &  \bf{Mission 1} &  \bf{Mission 2} \\ \hline
    ${m_0}$ (kg) &  $30000$ & $30000$ \\ \hline
    ${{\bm{r}}_{{{\cal I}}{\rm{0}}}}$ (m) & ${\left[ 200 \ 200\ 1500 \right]^T}$  & ${\left[ 0 \ 0\ 1500 \right]^T}$\\ \hline
    ${{\bm{v}}_{{{\cal I}}{\rm{0}}}}$ (m/s) & ${\left[ -20 \ -20\ -80 \right]^T}$ & ${\left[ 0 \ 0\ -80 \right]^T}$\\ \hline
    ${{\bm{q}}_0}$ & $Euler2Quater^{*}({\left[ -20 \ 20\ 0 \right]^T})$ & ${\left[ 0 \ 0\ 0\ 1 \right]^T}$ \\ \hline
    ${{\bm{\omega }}_{{{\cal B}}{\rm{0}}}}$ (deg/s) & ${\left[ 0 \ 0\ 0 \right]^T}$ & ${\left[ 0 \ 0\ 0 \right]^T}$ \\\hline
    ${\gamma _{c }}$ (deg)  & $45$ & $45$ \\\hline
    ${\vartheta _{\max }}$ (deg)  & $30$ & $30$ \\\hline
  \end{tabular}
  \begin{tablenotes} 
        \footnotesize 
        \item[*] Function that convert Euler angles (deg) to quaternions.
  \end{tablenotes} 
  \end{threeparttable} 
\end{table}

The computation time of Mission 1 and Mission 2 is 0.62 s and 0.94 s, respectively. In Mission 1, the DNN takes 0.11 s to generate the initial trajectory, and the SCP algorithm takes 0.51 s to solve the guidance problem after 2 iterations. In Mission 2, the DNN takes  0.17 s to generate the initial trajectory, and the SCP algorithm takes 0.77 s to solve the guidance problem after 3 iterations. For real-time landing guidance, it is required to solve the trajectory online in less than 1 s \cite{szmuk2018,szmuk2020,reynolds2020}. In work \cite{szmuk2020}, it was claimed that the propagation step (see Eq. \ref{Propagation}) consumes time on the order of 10 ms by using C++ and Eigen matrix library. Thus, the computation time of the propagation step was omitted in \cite{szmuk2020}. However, all of our work is implemented in Python in this paper, and the computation time of the propagation step is almost the same as that of the solver. To provide more complete test results, the computation time of the propagation step is not omitted in this paper.

\begin{figure*}[!hb]
\centering
\subfloat[]{\includegraphics[width=8cm]{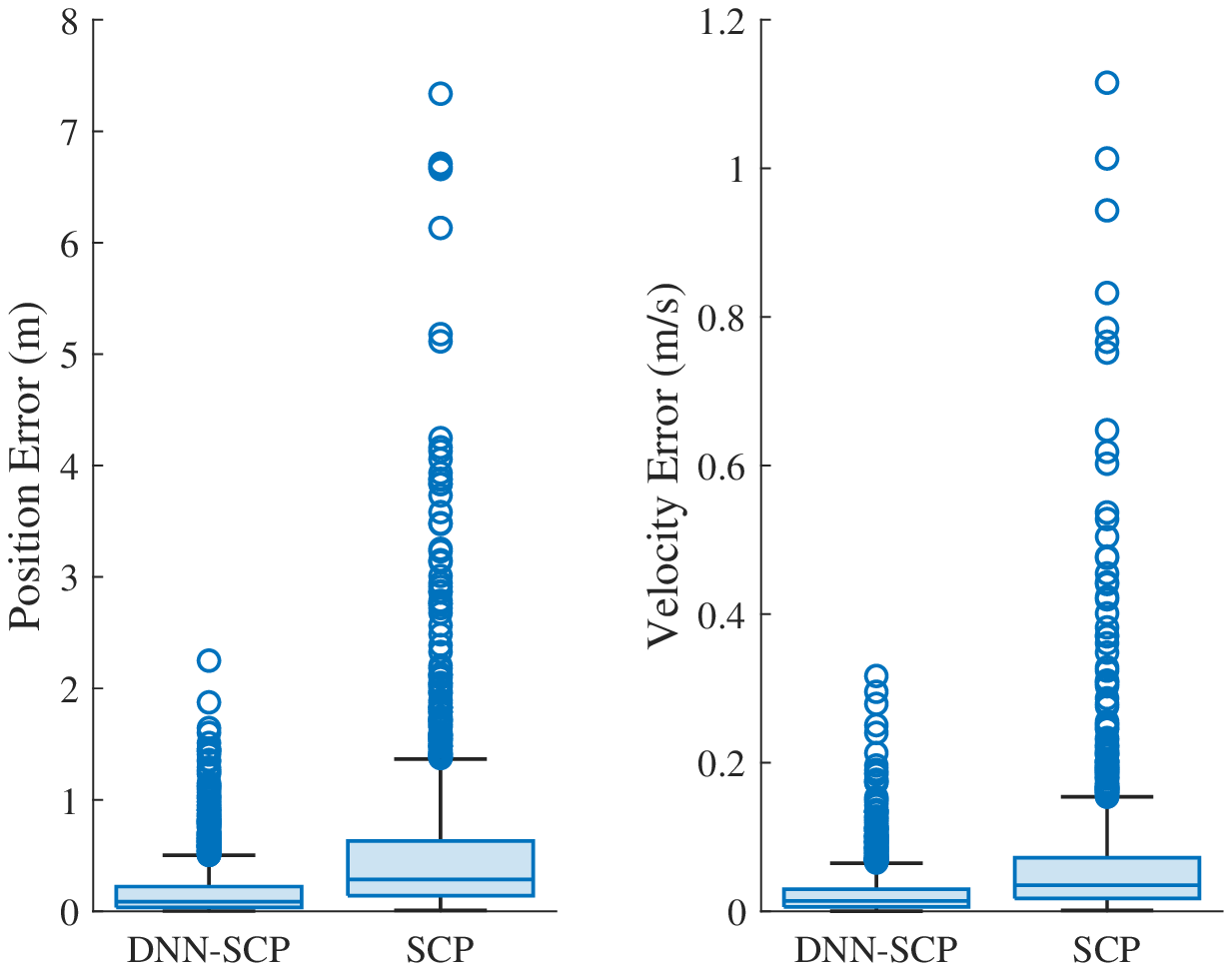}%
\label{fig7a}}
\hfil
\subfloat[]{\includegraphics[width=8cm]{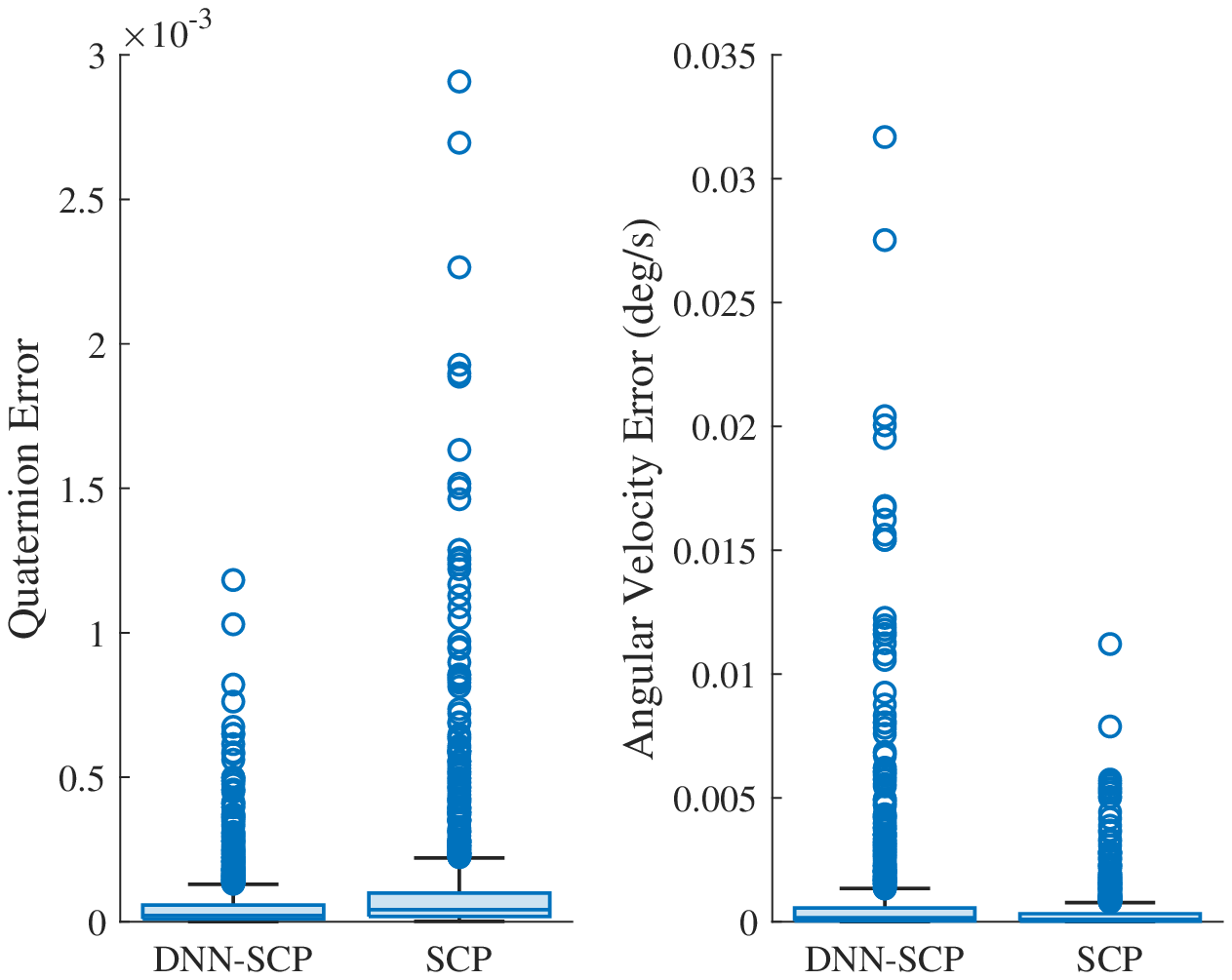}%
\label{fig7b}}
\caption{Errors of terminal states. (a) Terminal position and velocity errors. (b) Terminal orientation and angular velocity errors.}
\label{fig7}
\end{figure*}
\begin{figure*}[!ht]
\centering
\subfloat[]{\includegraphics[width=8cm]{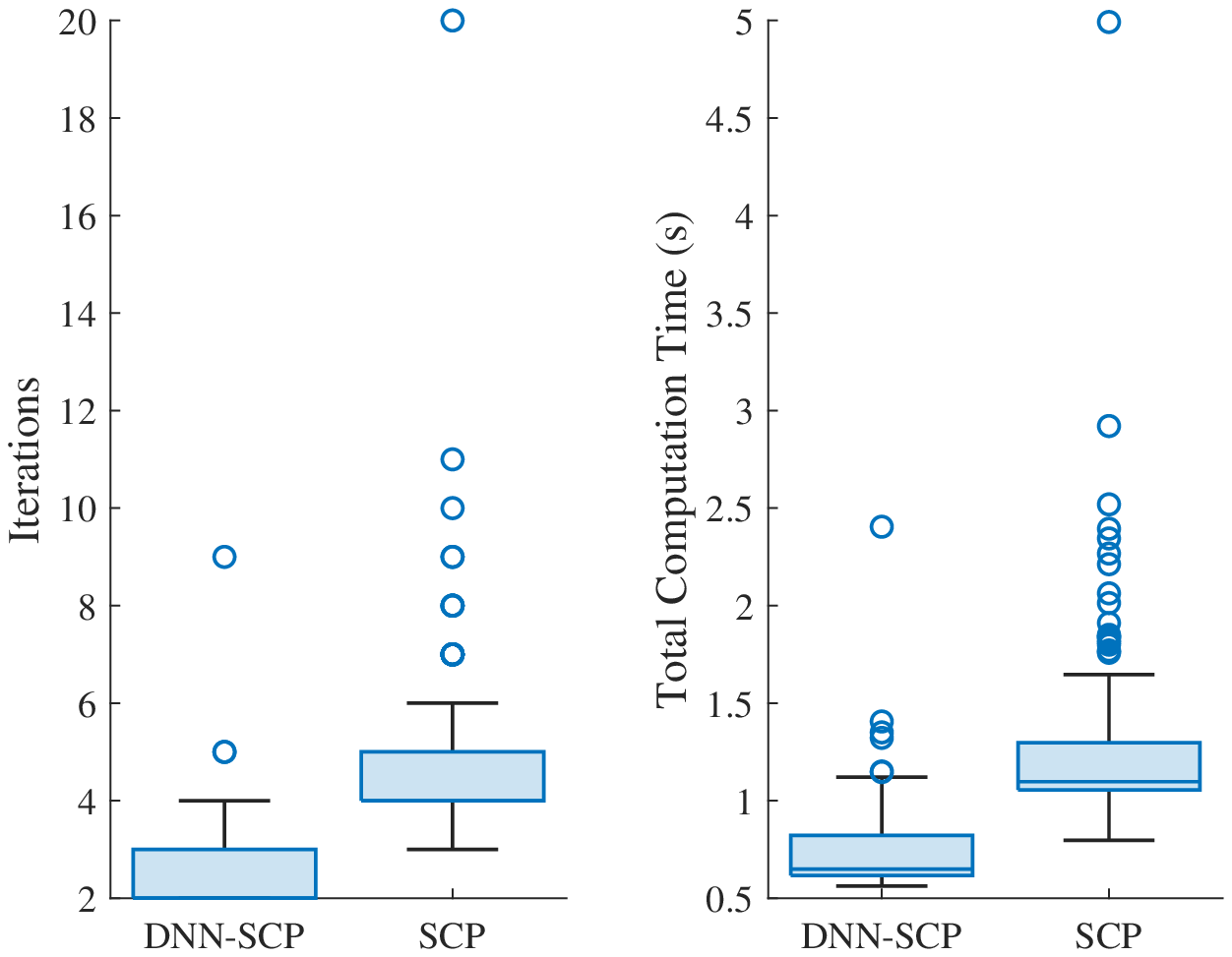}%
\label{fig8a}}
\hfil
\subfloat[]{\includegraphics[width=8cm]{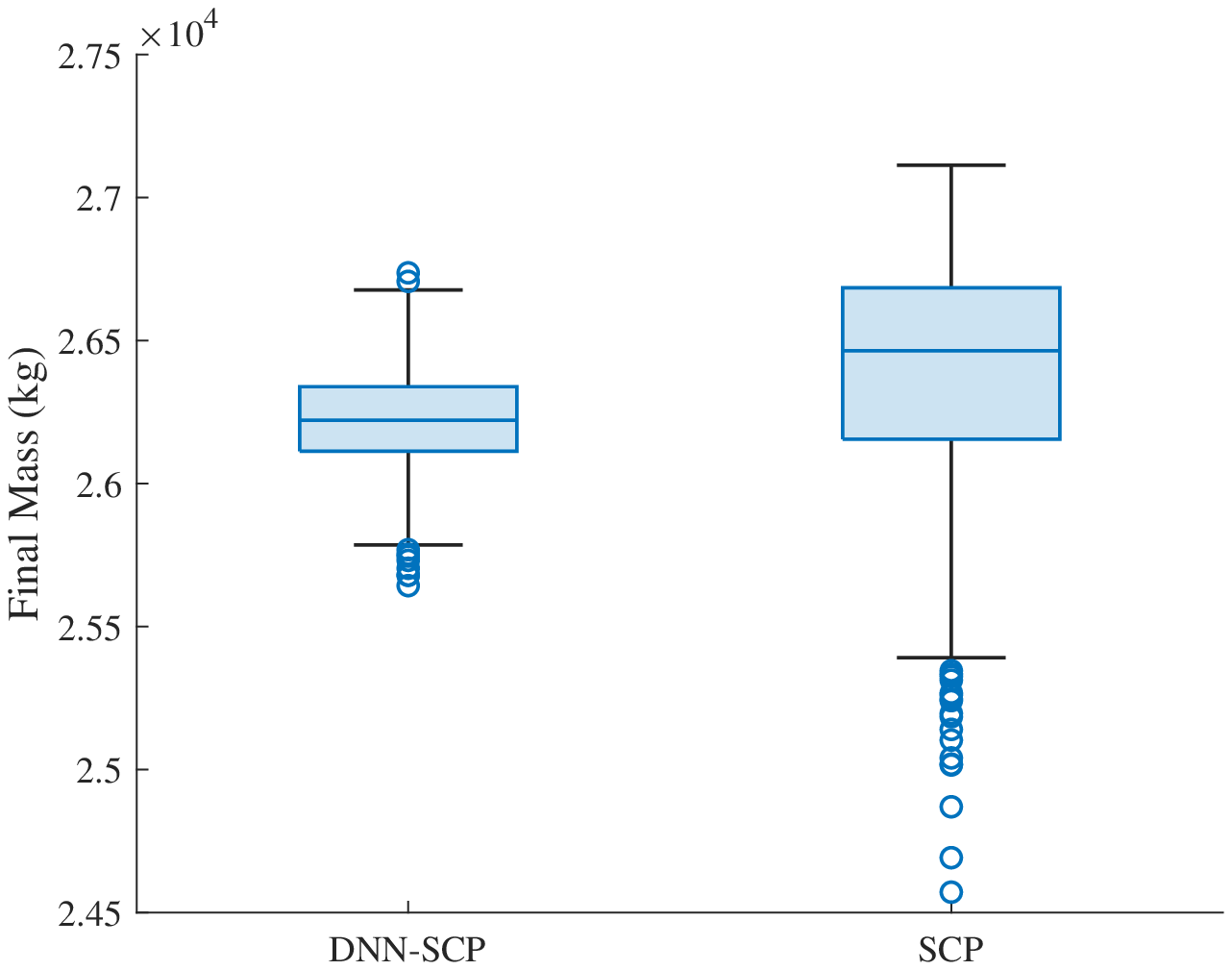}%
\label{fig8b}}
\caption{Computation performance and fuel remainder. (a) Iterations and total computation time. (b) Final mass. The higher the final mass is, the more fuel remains.}
\label{fig8}
\end{figure*}

To further analyze the performance of the proposed method, the Monte Carlo analysis is used to test it. The random parameter selection for Monte Carlo analysis is the same as that used to generate the data set in Sec. \ref{Dataset results}. In Monte Carlo analysis, the obtained thrust solution is used as the open-loop control command. The nonlinear dynamics of the vehicle are propagated using the open-loop control command. The proposed method is compared with the state-of-the-art SCP method. 1000 Monte Carlo simulations have been carried out, and Figs. \ref{fig7} and \ref{fig8} show the results. To simplify, the proposed method is labeled DNN-SCP in the Figs. \ref{fig7} and \ref{fig8}. The error is defined as the 2-norm of each state error, which can be expressed as
\begin{equation}\label{Error}
err({\bm{y}}) \buildrel \Delta \over = \left\| {{\bm{y}}({t_f}) - {{\bm{y}}_f}} \right\|
\end{equation}
where ${\bm{y}} \in \{ {{\bm{r}}_{{\cal I}}},{{\bm{v}}_{{\cal I}}},{{\bm{q}}_{{{\cal I}{\cal B}}}},{{\bm{\omega }}_{{\cal B}}}\} $, and ${{\bm{y}}_f}$ is the terminal constraint value of variable $\bm{y}$. The purpose of the landing guidance is to successfully land the vehicle on the landing site. Thus, the accuracy of the terminal states is significant. The errors of the terminal states are shown in Fig. \ref{fig7}. According to Fig. \ref{fig7a}, the proposed method performs better than the state-of-the-art SCP method in position accuracy and velocity accuracy. According to Fig. \ref{fig7b}, the proposed method is slightly better than the SCP method in orientation accuracy. And the angular velocity accuracy performance is almost the same. Even if the SCP method seems barely better than the proposed method in angular velocity accuracy, the angular velocity errors of the two methods are both small enough. The reason why the proposed method can obtain better accuracy is that, although the convergence criteria of the two methods are the same, the trained DNN provides a better initial trajectory, making it easier to obtain a dynamically feasible solution.

Another important performance for the proposed method is its computational performance. To meet the real-time requirements, the computation time needs to be less than 1 s \cite{szmuk2018,szmuk2020,reynolds2020}. According to Fig. \ref{fig8a}, the real-time performance of the proposed method is also better than that of the SCP method. Although a few tests do not meet the real-time requirements, most of the tests can meet the requirements. In this Monte Carlo simulation of 1000 tests, 991 tests meet the real-time requirements. And the median computation time is almost half of the SCP method. Specifically, the average computation time of the proposed method is 0.7035 s, and that of the SCP method is 1.1887 s, which is reduced by 40.8$\% $. Because the proposed method has a better initial trajectory provided by the DNN, it can converge faster. The median number of iterations of the proposed method is 2, while that of the SCP method is 4. Even if the DNN needs an average of 0.1523 s to generate the initial trajectory, the computation time of the proposed method is far less than that of the SCP method. As this paper studies the minimum-fuel guidance problem, Fig. \ref{fig8b} shows the final mass of the vehicle. The greater the mass remaining, the more fuel remaining. The proposed method generally consumes slightly more fuel than the SCP method. The trajectory of the proposed method depends largely on the initial trajectory generated by the DNN, so its optimality is also limited by the initial trajectory.

In general, a good initial trajectory based on historical experience can improve the performance of iterative algorithms such as the SCP algorithm. However, it is very inefficient to use historical experience data directly. A better initial trajectory generator can be obtained by training the neural network with historical experience data. The initial trajectory generator can make the algorithm converge faster by giving a satisfactory initial trajectory based on the training data. In addition, the trained neural network does not occupy much memory and has high computational efficiency.

\section{Conclusion}\label{Conclusion}
This paper proposes a real-time computational guidance method using data-driven initialization. We present a guidance method that combines the SCP method with neural networks in the framework of 6-DoF dynamics considering aerodynamic effects. Instead of brutally using neural networks as the controller, we use a DNN as an initial trajectory generator to generate the initial trajectory required by the SCP method. This makes the proposed method less time-consuming than the state-of-the-art SCP method. And compared with using neural networks as the controller, the SCP algorithm ensures reliability. In extensive Monte Carlo tests, the proposed approach outperforms a state-of-the-art SCP method. The terminal states accuracy of the proposed method is better. And the proposed method consumes less time, saving 40.8$\%$ computation time compared with the SCP method. 99.1$\%$ of the test cases take less than 1 s, which means that the proposed method is more suitable for online real-time applications. Furthermore, our DNN-based SCP scheme opens up interesting lines of future research. The DNN-based trajectory generator is designed as a sequence model predictor, which makes the proposed architecture can be used to improve the performance of the SCP algorithm in various applications.


 
%

\bibliographystyle{IEEEtran}

\bibliography{Reference2}

\end{document}